\documentclass[12pt,preprint]{aastex}
\slugcomment{To appear in ApJ}

\shorttitle{}
\shortauthors{Cieza et al.}

\begin{document}


\title{Testing the Disk Regulation Paradigm with Spitzer Observations. I. \\
Rotation Periods of Pre-main Sequence Stars in the IC 348 Cluster}


\author{Lucas Cieza and Nairn Baliber}
\email{lcieza@astro.as.utexas.edu, baliber@astro.as.utexas.edu}

\affil{Astronomy Department, University of Texas at Austin \\
      University Station C1400, Austin, TX 78712}

\begin{abstract}
We present 105 stellar rotation periods in the young cluster IC\,348,
75 of which are new detections, increasing the total number of known
periods in this cluster to 143. The period distribution resembles that
seen in the heart of the Orion Nebula Cluster by Herbst and
colleagues.  Stars estimated to be less massive than 0.25\,$M_{\odot}$
show a unimodal distribution of fast rotators (P\,$\sim$\,1--2~days) and 
a tail of slower rotators, while stars estimated to be more massive than 
0.25\,$M_{\odot}$ show a bimodal distribution with peaks at $\sim$\,2 and
 $\sim$\,8 days.  We combine all published rotation periods in IC\,348 with 
\emph{Spitzer} mid-IR (3.6, 4.5, 5.8, and 8.0\,\micron) photometry, which 
provides an unprecedented efficient and
reliable disk indicator in order to test the disk-braking paradigm. We
find no evidence that the tail of slow rotators in low-mass stars or
the long period peak in high-mass stars are preferentially populated
by objects with disks as might be expected based on the current
disk-braking model. Also, we find no significant correlation between
period and the \emph{magnitude} of the IR-excess, regardless of the
mass range considered. Given the significant improvement of
\emph{Spitzer} observations over near-IR indicators of inner disks,
our results \emph{do not} support a strong correlation in this cluster
between rotation period and the presence of a disk as predicted by
disk-braking theory.  Rather, they are consistent with the suggestion
that the correlation between period and the amplitude of the
(I\,--\,K) excess reported in the past is a secondary manifestation of
the correlation between the amplitude of near-IR excess and mass. By
comparing our sample with a recent \emph{Spitzer} census of IC\,348,
we find that the disk properties of our sample are indistinguishable
from the overall disk properties of the cluster. We conclude that it
is very unlikely that the lack of a correlation between rotation
period and IR excess is due to a bias in the disk properties of our
sample.  Finally, we find some indication that the disk fraction
decreases significantly for stars with very short periods (P\,$\lesssim$\,1.5
days). The fact that very fast rotators tend to have little or no
excess has already been recognized by Rebull and colleagues for stars
in the Orion Nebula Cluster.  This is the only feature of our sample
that could \emph{potentially} be interpreted as evidence for disk
braking.  There is currently no alternative model to disk braking to
explain the evolution of the angular momentum of pre-main-sequence
stars.  It has been proposed that the observational signatures of disk
braking might be significantly masked by the intrinsic breadth of the
initial period distribution.  We argue that more rigorous 
modeling of angular momentum evolution and a quantitative analysis of 
the observational data are required before the disk-braking model can 
be regarded as inconsistent with observations.

\end{abstract}


\section{Introduction}
The evolution of the angular momentum of pre-main-sequence (PMS) stars
is one of the longest-standing problems in star formation and is
currently a controversial topic. As low-mass
($\sim$0.1--1.2\,$M_{\odot}$) PMS stars evolve along their convective
tracks, conservation of angular momentum dictates that they should
spin up considerably.  Low-mass PMS stars are believed to contract by
a factor of $\sim$2--3 during their first 3\,Myrs of evolution.  If
these stars were to conserve their specific angular momentum, \emph{j}
(where \emph{j}\,$\propto$\,$R^{2}/P$), their rotation periods would
be expected to decrease by a factor of $\sim$\,4--9. Using the models
by \citet[D98, hereafter]{dantona94, dantona98},\footnote{Available at
http://www.mporzio.astro.it/~dantona/prems.html} and assuming angular
momentum conservation and homologous contraction, \citet{herbst00}
show that, given a starting period of 10 days,\footnote{Adopting an
initial period of 10 days is a conservative choice because recent
observations \citep[e.g.][]{covey05} show that deeply embedded PMS
stars tend to rotate \emph{faster} than optically revealed PMS stars.}
all low-mass stars should rotate with periods shorter than $\sim$2
days by an age of 2\,Myrs.  However, observations show a large
population of slow rotators (P\,$\sim$\,8.0 days) in clusters that,
according to the same evolutionary tracks, are $\sim$2\,Myrs old or
older \citep{luhman03}. Also, the \emph{very} broad distribution of
rotation periods in the zero-age main sequence (ZAMS) suggests a very
broad range of angular momentum loss between the birth line and the
ZAMS.

Proposed mechanisms to explain the presence of slow rotators come in
the form of magnetic-field interaction between the PMS star and
material in its inner disk, locking the star's rotation to the
Keplerian velocity of the disk's inner edge \citep{konigl91, shu94}
and/or creating an accretion-driven wind \citep{shu00}, thereby
transferring angular momentum from the star to material from the disk,
preventing it from spinning up, unchecked.  ``Disk locking'' and
``disk braking'' have become essential features of virtually every
model of the angular momentum evolution of PMS stars.

The first order prediction of disk-locking and disk-braking models is
that, as a group, stars with disks should rotate slower than stars
that have already lost their disks and are free to spin up.  However,
finding an observational correlation between the presence (or lack) of
a circumstellar disk and the rotation period of PMS stars has proven a
difficult task. It has been repeatedly argued that one of the main
difficulties in finding this predicted correlation is the lack of an
appropriate disk indicator \citep{rebull01, herbst02}.
\emph{Spitzer}'s unprecedented mid-IR sensitivity allows, for the
first time, an unambiguous identification of disks in a
statistically significant sample of PMS stars with known rotation
periods.

Here we report 75 previously unidentified rotation periods detected in
the young cluster IC\,348 which is situated in the East side of the
Perseus Molecular Cloud. Our new periods, when combined with the
periods in the literature, increase the total number of known periods
in this cluster to 143. IC\,348 is 315\,pc away, young (2--3 Myrs),
relatively compact ($\sim$400 members, D\,$\sim$\,20 arcminutes), and
has low extinction (A$_{{\rm V}}$\,$<$\,4) \citep{luhman03}. We combine all
published rotation periods in IC\,348 with \emph{Spitzer} photometry
to investigate the disk-braking scenario. This paper is organized as
follows. In Section~\ref{previous}, we summarize previous
observational attempts to identify a period--disk correlation.  In
Section~\ref{photometry}, we describe our time series photometry and
the observed rotation period distribution.  In Section~\ref{rotation},
we combine the rotation periods with the disk identification from
\emph{Spitzer} observations in order to search for the period--disk
correlation predicted by the disk-braking paradigm.  In section 5, we
discuss the implication of our results in the context of the
disk-braking scenario. Finally, we summarize our results in Section 6.

\section{Previous Observational Results}\label{previous}
\subsection{Rotation Periods and Disks in Other Clusters}
Rotation periods of young stars with late spectral types can
efficiently be obtained via precise differential photometry due the
brightness modulation produced by the rotation of cool star spots and
hot accretion columns reaching their surfaces.  
Fifteen years ago, the number of PMS stars with
photometrically measured rotation periods was less than 100. This
number is currently approaching 2000 (see \citet{stassun03} for a
recent review of observations). As the number of available rotation
periods of PMS stars in young stellar clusters has increased, some
groups have found apparent observational signatures of the
disk-locking model. \cite{attridge92} report a bimodal distribution in
the period of 35 PMS stars in the Orion Nebula Cluster (ONC), peaking
at $\sim$2 and $\sim$8 days. This bimodal distribution has been
confirmed by \citet[H00 and H02, hereafter]{herbst00, herbst02}, who
find it to be restricted to samples of stars with masses
$>$~0.25\,$M_{\odot}$. H02 find that very low-mass stars
(M~$<$~0.25\,$M_{\odot}$) exhibit a unimodal period distribution
dominated by fast rotators with periods of 1--2 days. H00 and H02 also
report a positive correlation between the presence of a circumstellar
disk (as indicated by the presence of K-band excess) and slow
rotation.  The long-period peak in the bimodal distribution has been
interpreted by these authors as being populated by disk-locked stars.

However, observations by several other groups do not confirm the
bimodal nature of the period distribution of PMS stars in the ONC.
Based on a sample of 254 stars, \citet{stassun99} find that the
distribution of rotation periods between 0.5 and 8.0 days is
statistically consistent with a constant distribution and that there
is no apparent correlation between rotation period and near-IR excess.
H00 argue that the \citet{stassun99} results do not show the expected
bimodal distribution because their sample is dominated by stars of
very low mass (i.e. M~$<$~0.25\,$M_{\odot}$). Still, based on 281
periods of stars in four fields \emph{around} (but not including) the
heart of the ONC, \citet{rebull01} finds that the distribution of
periods is statistically indistinguishable for stars less and more
massive than 0.25\,$M_{\odot}$.  \citet{herbst05} argue that the
``Flanking Fields'' observed by \citet{rebull01} represent a very
heterogeneous sample in terms of age in which any structure in the
period distribution would be wiped out.  \citet{rebull01} also finds
no clear correlation between the presence of a circumstellar disk and
three different disk indicators (I$_{{\rm C}}$--K, H--K, and U--V
color excesses). \citet{rebull01} and Hillenbrand et al. (1998) 
show that the correlation between near-IR excess and disk
presence is far from perfect, leading to many disks being missed and 
false identification of disks.  \citet{rebull01} does conclude, 
however, that, given the lack of any observable correlation, disk locking 
is probably not the complete solution to the period distribution of PMS 
stars. 

More recently, a statistically significant number of rotation periods
in NGC\,2264 have become available. In this cluster, \citet{lamm05}
find a bimodal distribution for a sample of 184 stars with estimated
masses $>$\,0.25\,$M_{\odot}$. They find that the peaks in the period
distribution observed in NGC\,2264 are shifted toward shorter 
periods with respect to the ones found in the ONC. These shifts are 
interpreted as evidence of angular momentum evolution between the
age of the ONC ($\sim$\,1 Myrs) and that of NGC 2264 ($\sim$\,2--4 Myrs).
Using R\,--\,H$\alpha$ color criteria for disk identification, they 
argue that stars with disks tend to
rotate slower than stars without disks.  However, 60$\%$ of their
sample have ambiguous disk identification. \citet{makidon04} report
periods for 118 stars in NGC\,2264 with estimated masses
$>$~0.25\,$M_{\odot}$. They find no evidence for a
statistically significant bimodal distribution or a disk--period
correlation, regardless of the disk indicator used: EW(H$\alpha$),
U--V, I--K, or H--K color excesses. Clearly, despite the categoric
claims made by some authors \citep[e.g.][]{herbst05}, both the
existence of a bimodal distribution in the rotation periods of PMS
stars and the presence of a disk--rotation period correlation still
await independent confirmation.

\subsection{Rotation Periods and Disks in IC\,348}

All of the currently known periods for IC\,348 can be collected from
three papers (prior to this one): \citet{cohen04},
\citet{littlefair05}, and \citet{kiziloglu05}. Based on an observing
campaign spanning more than 5 years, \citet{cohen04} report the
rotation periods of 28 PMS stars in IC\,348.  The observations were
taken in the $\rm{I_{C}}$ band with the 0.6\,m telescope at the Van
Vleck Observatory, which has a field of view of 10$\farcm$2 on a
side. \citet{littlefair05} report 32 additional periods based on
$\rm{I_{C}}$ observations performed with the 1.0\,m Jacobus Katelyn
Telescope, which has a usable field of view of 9$\arcmin$ on a
side. \citet{cohen04} and \citet{littlefair05} only report 18 periods
in common (partially because the \citet{littlefair05} observations are
considerably deeper than those presented by \citet{cohen04}), but the
periods in common show excellent agreement.  Finally,
\citet{kiziloglu05} present 35 periods based on observations taken
with the 0.45\,m ROTSE-IIId robotic telescope which has a field of
view of 1.85 degrees on a side and operates without a filter. Of these
35 periods, 17 were previously detected by \citet{cohen04}, one was
reported by \citet{littlefair05}, and 18 were new
detections.  \citet{kiziloglu05} calculated the confidence
level of their rotation periods assuming white noise as the only source
of error; however, as we discuss in Section\ref{findperiods}, that is
a very optimistic assumption. Therefore, we only use the rotation
periods from their work which have confidence levels greater than
5\,$\sigma$. 

\citet{cohen04} do not have enough stars to fully explore the presence
of a bimodal distribution or the disk--period correlation in their
sample but argue that the period distribution hints at the bimodality
seen in the ONC.  \citet{littlefair05} also see a hint of a period
bimodality in the ``high'' mass stars but argue that it is not
statistically significant given the small number of stars in the
appropriate mass range. They did explore the possibility of an
observable disk--period correlation in their sample. However, they
find no correlation of period with K\,--\,L color excess (available
for 30 stars) or ${\rm H}{\alpha}$ equivalent width (available for 43
stars).  Also, \citet{littlefair05} suggest that the rotation
period\,--\,(I--K) excess correlation reported by H02 might be due to
a secondary correlation arising from the fact that (I--K) excess is
easier to detect in stars more massive than 0.25\,$M_{\odot}$ than it
is to detect in lower-mass stars \citep{hillenbrand98}. The
\citet{kiziloglu05} paper focuses on an X-ray luminosity--period
correlation and does not discuss the shape of the period distribution
or any possible correlation between period and disk indicators.

\section{Optical Observations and Data Reduction}\label{photometry}

\subsection{Time Series Photometry}
For our photometric measurements of IC\,348 stellar rotation periods,
data were collected using the McDonald Observatory 0.76~m Telescope
and its Prime Focus Corrector (PFC) \citep{claver92}, which provides a
1$\degr$ field of view, 46$\farcm$2 x 46$\farcm$2 of which is covered
by a 2048 x 2048 CCD.  ${\rm I_{C}}$-band time-series photometry was
performed on data collected during 3 observing runs: 2003, December
6th--22nd; 2004, January 4th--8th; and 2004, January 24th--30th.  Ten
slightly overlapping fields were used to cover the entire IRAC map of
the Perseus Molecular Cloud obtained as part of the \emph{Spitzer}
Legacy Project ``From Molecular Cores to Planet-forming Disks,'' (c2d)
\citep{evans03}.  Two of these fields are centered on the IC\,348 and
NGC\,1333 clusters and are referred as `cluster fields' in the
following discussion.  Three frames were taken for every pointing of
the cluster fields: a 15 sec exposure followed by two 150 sec
exposures. For the rest of the fields, only two frames were taken,
namely, a 3 sec exposure followed by a 15 sec exposure. From the 3
observing runs, a total of $\sim$140 useful data points were obtained
for each star of the cluster fields and $\sim$50 data points for stars
in the rest of the fields.  Typical rotation periods of PMS stars
range from $\sim$0.2 to $\sim$15 days. 

We tested the range of periods recoverable in our data by inserting
sine waves of several different amplitudes into data from non-varying
stars of various magnitudes in the IC\,348 field.  The model
variations ranged in period from 0.3 to about 19 days.  Using the same
detection method in identifying rotation periods in IC\,348 PMS stars,
recovery rates for the model variations were about 71\% for periods
between 0.4 and 0.5 days and 52\% for periods between 0.5 and 0.6
days, although most of the periods that were not accurately identified
in this regime were a factor of two away from the actual period, which
would not affect the overall rotation period distribution
significantly.  Nearly 100\% of the models with periods ranging from
0.6 days and above were correctly identified.  Therefore, we conclude
that the observations obtained provide both an appropriate sampling
rate and baseline to be sensitive to all expected rotation periods.

After standard CCD reductions were performed with the IRAF
imred.ccdred package, photometry is performed using a combination of
the ISIS image subtraction software \citep{alard98,alard00} and our
own code.  After residual images are generated by ISIS, our code
measures aperture sums for stars in the residual images and produces
light curves measured in percent flux normalized to a reference image.
Here we present the rotation periods found in the field covering the
IC\,348 cluster.  Rotation periods for stars in the rest of the
Perseus cloud, including those in the cluster NGC\,1333, will be
presented in a future paper.

\subsection{Finding the Rotation Periods}\label{findperiods}
To search for periodic signals in the light curves of our targets, we
used the standard periodogram technique discussed by \citet{scargle82}
for analyzing periodic signals of unevenly spaced data and the
prescription given by \citet{horne86} to select the optimum number of
independent frequencies used to inspect the data.  The initial period
range of the periodogram was set to 0.1--50 days, but, since the number
of independent frequencies inspected is heavily weighted toward higher
frequencies, low frequencies are not well sampled in this range, which
produces a large uncertainty for long periods. For this reason, a new
periodogram was produced for every star.  In this new periodogram, the
shortest period sampled was set to 0.75 times the period found in the
first pass.  For our data set, the normalized power spectrum (PS) peak
of the Scargle periodogram corresponding to a false alarm probability
\citep[FAP, as defined by][]{horne86} of 1$\%$ is $\sim$10. This
threshold corresponds to the 1\% probability that a power spectrum
peak reaches a given high by pure fluctuations of white, uncorrelated
noise, and is clearly too optimistic. 

\begin{figure}
%\figurenum{1}
\epsscale{1}
\plotone{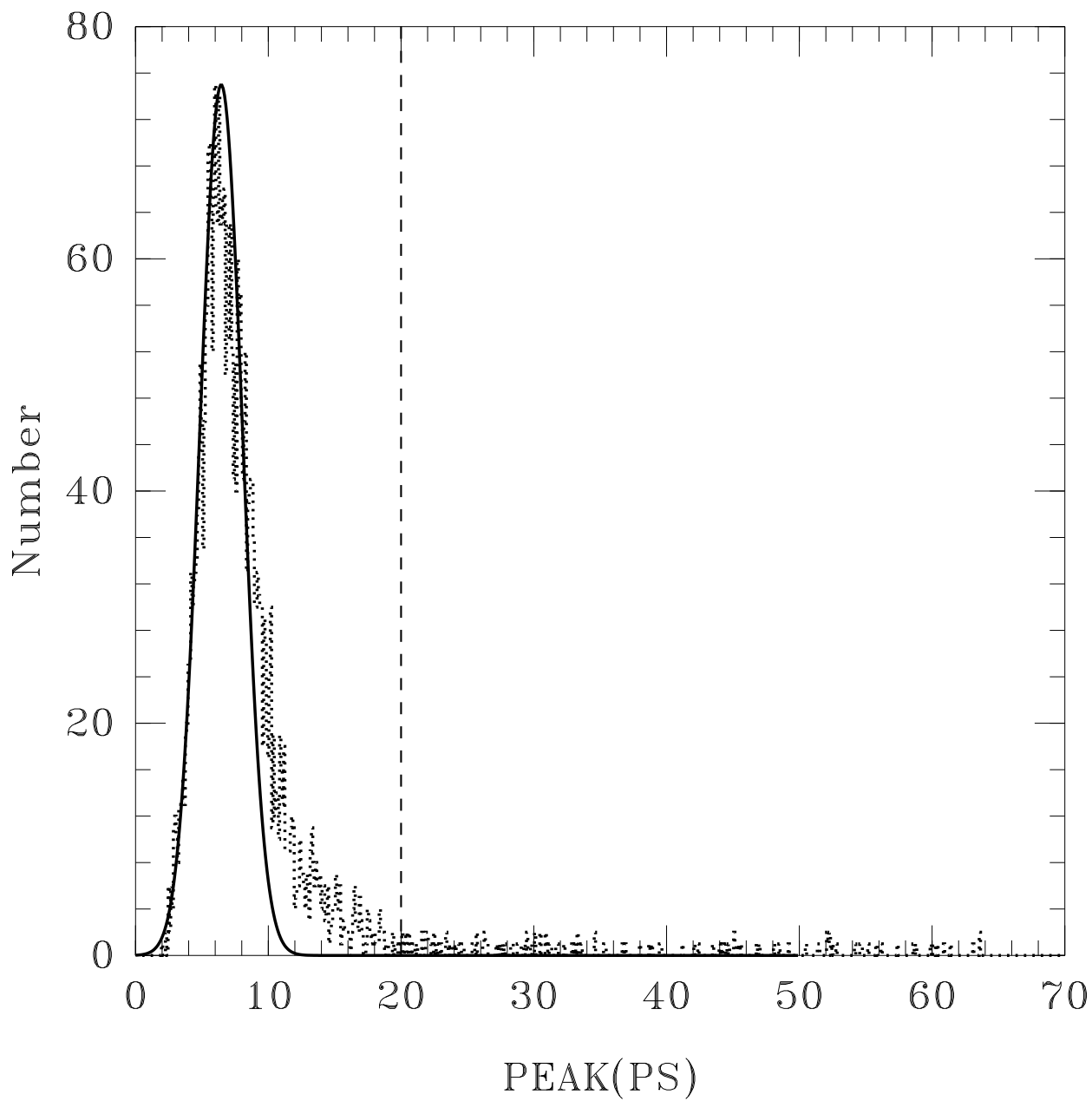}
\caption{This is a plot of the distribution of power spectrum peaks
for the entire sample of stars in the IC\,348 data set.  The dotted
line is a histogram of the maximum power spectrum peak of each star,
and the solid line is a Gaussian fit to the left side of the
distribution of power spectrum peaks. Non-white noise is likely to raise a 
particular peak above what white noise alone would provide.  Such systematic noise 
creates the shoulder above the Gaussian curve on the right wing of the distribution of
peaks.  This additional non-white noise makes period determinations
for stars with power spectrum peaks lower than $\sim$20
unreliable.\label{pspeakdist}}
\end{figure}

\begin{figure}
\plotone{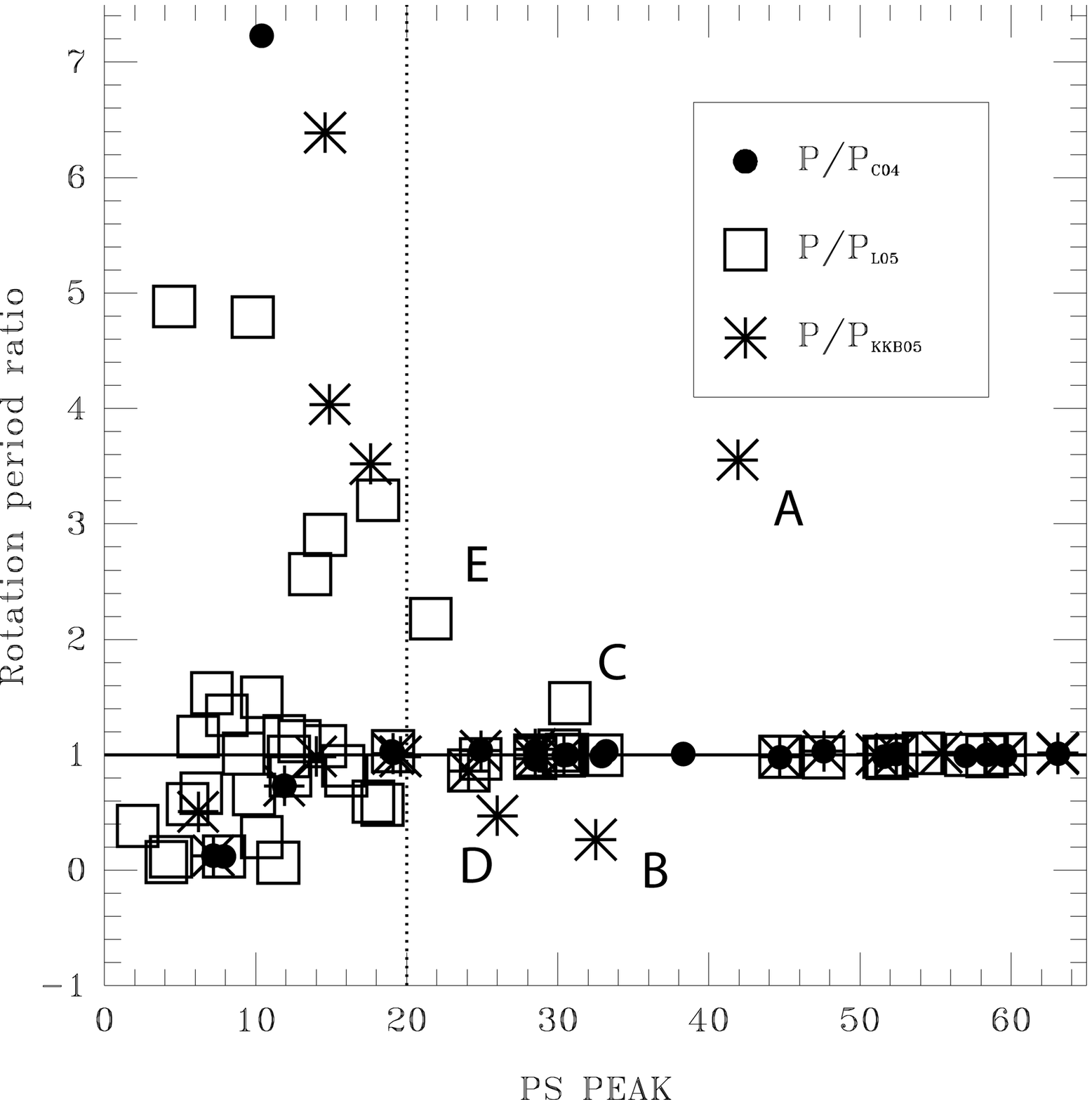}
\caption{This is a plot of the ratio of our rotation periods to those
from \citet[C04]{cohen04}, \citet[L05]{littlefair05}, and
\citet[KKB05]{kiziloglu05} as a function of power spectrum peak from
our light curves.  The values start to diverge for power spectra peaks
lower than $\sim$20.  Based on the peak distribution from
Figure~\ref{pspeakdist} and these ratios, we adopt a power spectrum
peak of 22.0 as the threshold above which we consider the measured
periodic signal to be an accurate representation of the stellar
rotation period.  The disagreement with rotation periods from the
literature for stars A through E is discussed in the text and in
Figure~\ref{disagreement}.\label{periodratios}}
\end{figure}

The assumption that the data points are statistically uncorrelated and
that noise can be characterized as white is usually not valid for time
series photometry of PMS stars \citep[see][and references
therein]{rebull01}. The distribution of PS peaks for the entire sample
of $\sim$3600 stars in the IC\,348 field is shown in
Figure~\ref{pspeakdist}. The left side of the peak can be
characterized as a Gaussian centered at 6.4 with a FWHM of
1.5 days. Rightward of the peak, the PS distribution has a tail
corresponding to light curves with real periodic signals combined with
spurious signals produced by non-white noise.  The shape of the
distribution of the PS peaks suggests that periods become unreliable
for PS peaks lower than $\sim$20.  Since our field contains over 60
stars with known rotation periods from the literature, we can also
estimate our PS confidence threshold by plotting the agreement of our
periods with published values from \citet{cohen04},
\citet{littlefair05}, and \citet{kiziloglu05} as a function of the
peak of the PS (Figure~\ref{periodratios}).

\begin{figure}
%\figurenum{3}
\epsscale{0.7}
\plotone{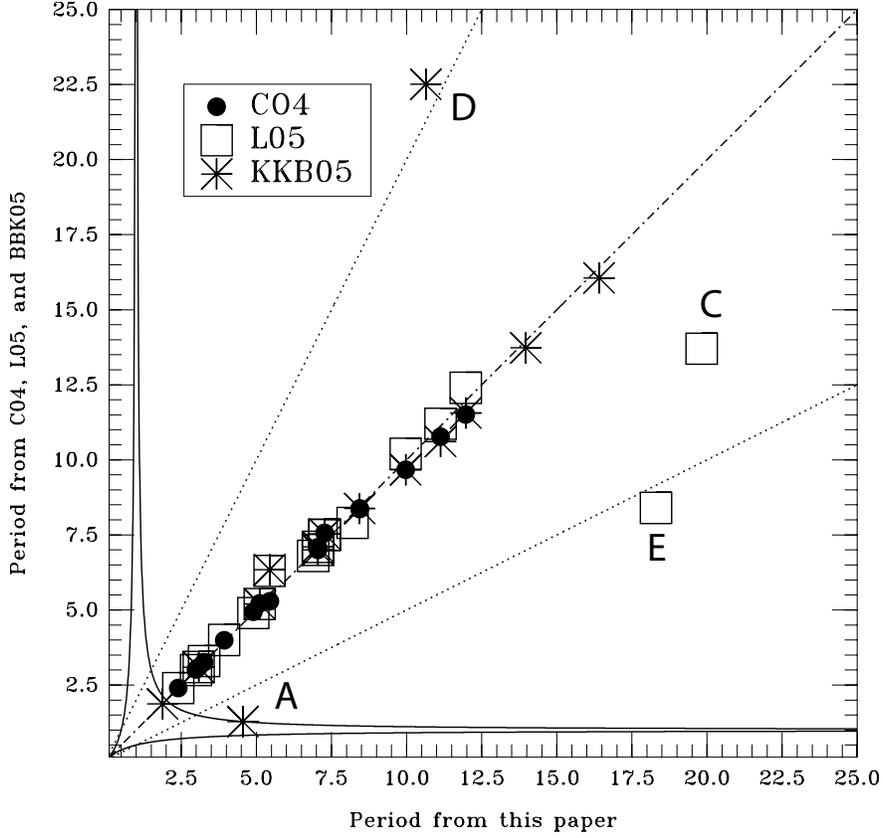}
\caption{This figure is another comparison of our IC\,348 periods to
those obtained by other groups.  Only five stars for which we measure
power spectra peaks above 20 show significant discrepancies between
our periods and those from the literature.  For star A,
\citet[KKB05]{kiziloglu05} find a period of 1.28 days, where we
measure a period of 4.55 days.  Their shorter period is a result of
the beating of our period at a 1-day sampling interval, of the form
$1/{\rm P_{beating}}$~=~$\pm$\,1~$\pm$~(1/P) (plotted as a solid line).  For
object B and D, we find shorter periods than those found by KKB05;
their periods are likely harmonics of the real periods (the two dotted
lines represent factor-of-two harmonics).  Our period for object C,
19.8 days, shows a discrepancy with the 13.4-day period found by
\citet[L05]{littlefair05} which cannot easily be explained.  Longer
periods have greater uncertainty, however, and since our observational
baseline is twice as long as that of L05 (52 {\it versus} 26 days), we take
our period to be closer to the correct value.  Our period for object E
is 18.3 days while L05 measured a period of 8.4 days. In this case, our
period is likely a harmonic of the real period.\label{disagreement}}
\end{figure}

Figure~\ref{periodratios} shows that even though some of our periods
agree with published values down to a PS peak of $\sim$10, many values
start to diverge when the peak is $\sim$20.  Above a PS peak of 20,
the periods we find disagree with published periods for only 5
objects, labeled A through E in Figures~\ref{periodratios} and
\ref{disagreement}.  In what follows we analyze these objects one by
one.  Object A (ID 91) has a PS peak of 41.9. We find a period of 
4.55 days while \citet{kiziloglu05} finds a period of 1.28 days. 
Figure~\ref{disagreement} shows that the period found by
\citet{kiziloglu05} can be explained as the beating of our period and
a 1 day sampling interval (i.e. a relation of the form
$1/{\rm P_{beating}}$~=~$\pm$\,1~$\pm$~(1/P), denoted by the solid
line). Object B (ID 102) has a PS peak of 32.5, and we find a rotation 
period of 8.57 days. \citet{kiziloglu05} assign to this object a rotation
period of 32.28 days, which is much longer that the typical rotation
periods of PMS stars.  Since the period found by them is close to a
factor of four larger than our period, it is likely to be an harmonic
of the real period.

For object C (ID 98), which has a PS peak of 30.8, we find a period
of 19.8 days, while \citet{littlefair05} find a period of 13.4
days. Both cases represent an unusually long rotation period and the
discrepancy is likely to be due to the increasing uncertainty with
increasing period for relatively short observing campaigns. Since our
observations span 52 days and \citet{littlefair05} observations span
only 26 days, our period is more likely to be correct.  For object D
(ID 123), we find a PS peak of 26.0 and a period of 10.64 days.
\citet{kiziloglu05} find a period of 22.51 days, which is also likely
to be an harmonic of the real period.  Finally, for object E, we find
a PS peak of 21.6 and a period of 18.3 days.  \citet{littlefair05}
find a period of 8.4 days. In this case, it is likely that our period
is an harmonic of the real period found by the other group.  The
contamination of rotation-period distributions by harmonics and ``beat
periods'' at the 10$\%$ level has already been seen by many groups
\citep[e.g. H02 and][]{lamm05} when comparing their results to
previously published data.  In addition to a change in the beating of
a period within the sampling rate (which would be different for each
observing run), the observed periods themselves may change between
observing runs \citep[as discussed in][]{rebull01} due to the
migration of stellar spots in a differentially rotating
photosphere. Clearly, the lower the adopted PS threshold, the larger
the contamination fraction. For definitiveness, based on
Figures~\ref{pspeakdist} through \ref{disagreement} and the above
discussion, we adopt a PS peak of 22.0 as the criteria to consider the
periodic signal to be an accurate representation of the rotation
period.  This corresponds to the highest PS peak of an object showing
a significant discrepancy with published periods that we believe is
due to our own incorrect period (object E in Figures
\ref{periodratios} and \ref{disagreement}).

\subsection{The Rotation Period Distribution}
\begin{figure}
%\figurenum{4}
\epsscale{1}
\plotone{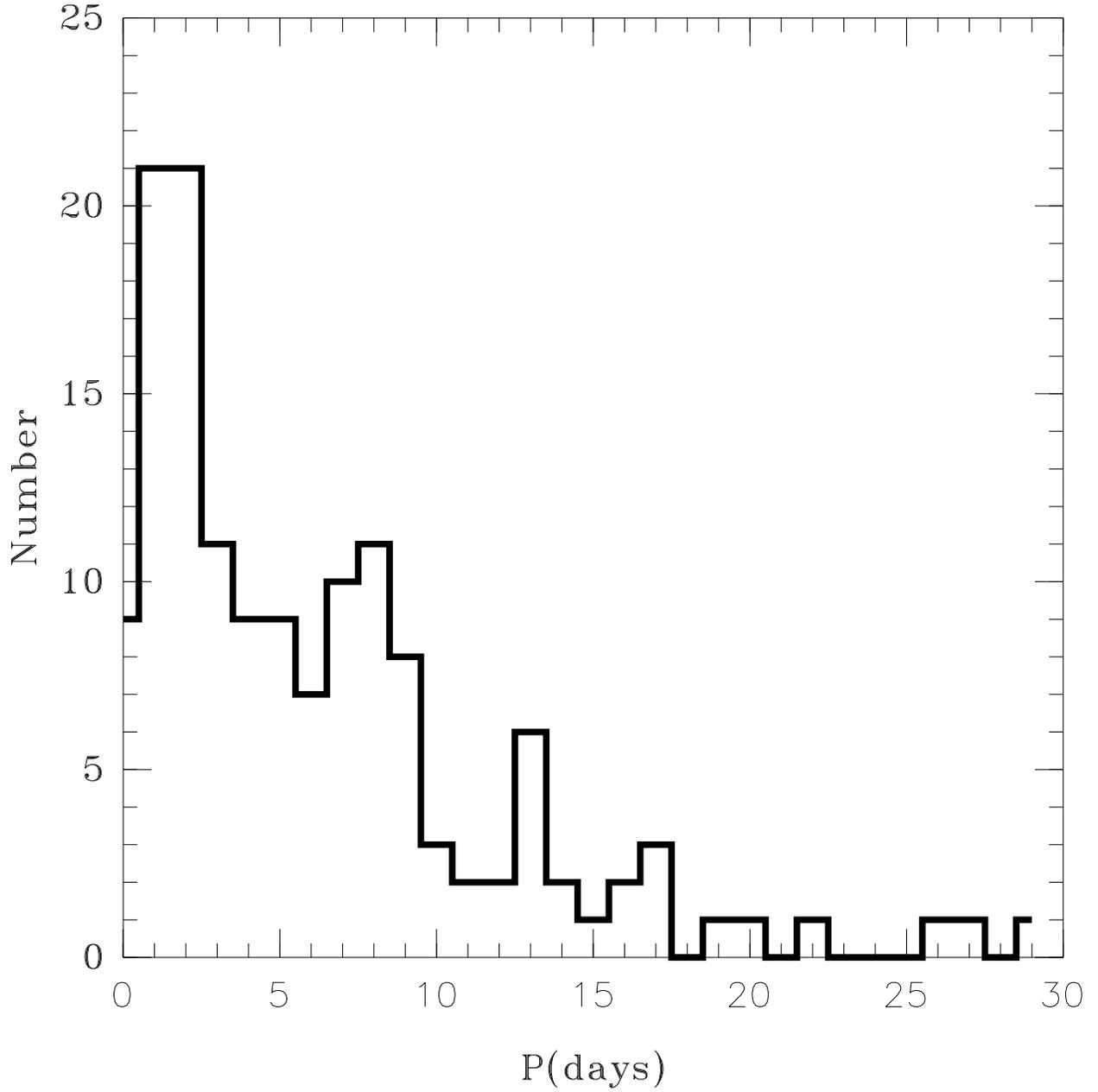}
\caption{A histogram of all 143 know rotation periods in
IC\,348 from our data and the literature.\label{histall}}
\end{figure}

Inspecting the periodograms of the $\sim$3600 stars detected in our
field, we find 105 objects with PS peaks higher than 22, the adopted
detection threshold. Of these 105 objects, only 30 had previously
known rotation periods. Thus, we have obtained 75 new rotation
periods.  Our periods more than double the number of previously known
rotation periods in the IC\,348 cluster and increase the total number
to 143; a comprehensive list of all known stars with periods in
IC\,348 appears in Table~\ref{xgfp_parameters}.  The coordinates of
each periodic stars listed are those of the closest counterpart
identified in the 2MASS survey.  Figure~\ref{histall} shows the
histogram of all the rotation periods listed in
Table~\ref{xgfp_parameters}. The period distribution extends to almost
30 days; however, for the rest of the paper, we restrict our analysis
to the 132 stars listed in Table~\ref{xgfp_parameters} with periods
shorter than 15 days. We decided to make this cut because stars with
periods $>$\,15 days ($\sim$8$\%$ of the total sample) are likely
to be more contaminated by harmonics than the rest of of stars (in the
previous section we found that periods longer than 15 days account for
3 out the 5 stars with PS peaks larger than 22 which have a period
disagreement with previously published values).  Also, any MS star
contaminating our sample is likely to have a rotation period $>$\,15
days.

\subsection{The Periodic Sample}\label{oursample}
The stellar population of IC\,348 and its disk properties are very
well characterized. \citet{luhman03} present a spectroscopic census of
the the cluster members.  This census is nearly complete 
within the central 16$\arcmin$\,x\,14$\arcmin$ area of the cluster for 
stars with spectral types M8 and earlier and yields a sample of 288 
objects whose membership has been established based on their proper motions,
positions in the H--R diagram, ${\rm A_{V}}$'s, and spectroscopic
signatures of youth. More recently, \citet{lada06} present
\emph{Spitzer} observations of all the 288 objects identified by
\citet{luhman03} as IC\,348 members. These results are based on the
the same IRAC GTO observations we use for disk identification, and
will be discussed further in Section~\ref{sampbias}.

Of the 143 periodic stars listed in Table 1, 92 are identified by
\citet{luhman03} as members of the cluster, and each of them have
measured spectral types.  We note that the 51 stars in our sample
which are \emph{not} identified by \citet{luhman03} as IC\,348 members
fall outside the 16$\arcmin$\,x\,14$\arcmin$ field for which they
present a complete membership census. However, \citet{cambresy06}
present an extinction map of the IC\,348 region and conclude that the
cluster extends up to 25$\arcmin$ from its center, placing all the
objects in our sample within those cluster boundaries.  Most of the
periodic stars in our sample turned out to be known members of
IC\,348, while only 8$\%$ of all of the stellar light curves we
inspected correspond to the 288 known members identified by
\citet{luhman03}. More importantly, none of our periodic stars 
correspond to any of the 123 targets identified by \citet{luhman03} 
as foreground or background non-members
of the cluster.  This indicates that periodicity is a very efficient
selector for PMS stars, and we conclude that it is therefore likely
that most, if not all, of the periodic stars falling outside the
region studied by \citet{luhman03} are in fact members of the IC\,348
cluster.

\subsection{Mid-IR Observations and Data Reduction}

As part of the c2d Legacy Project (Program ID 178), \emph{Spitzer} has
mapped 3.8 sq. deg. of the Perseus Molecular Cloud with the Infrared
Array Camera (IRAC, 3.6, 4.5, 5.8, and 8.0\,\micron) containing the
IC\,348 cluster and its surroundings. The IRAC maps consist of four
dithers of 10.4  sec observations divided into two epochs (i.e. a total
of 41.6\,sec exposures per pixel) separated by several hours. The
second-epoch observations were taken in the High Dynamic Range mode,
which includes 0.4\,sec observations before the 10.4 sec exposures,
allowing photometry of both bright and faint stars at the same
time. See \citet{jorgensen06} for a detailed
discussion of the c2d IRAC observations of the Perseus Molecular
Cloud. Also, \emph{Spitzer} has obtained deep observations IC\,348 as
part of the Guaranteed Time Observer program (GTO) ``Deep IRAC imaging
of Brown Dwarfs in Star Forming Clusters'' (Program ID 36). These
observations cover a 15$\arcmin$\,x\,15$\arcmin$ field of view
centered in the cluster and consists of two pairs of 8 dithers of 96.8
sec exposures for the 3.6, 4.5, and 5.8\,\micron\ observations
(1549 sec exposures per pixel). A characteristic of this
observing mode splits each $\sim$100\,sec IRAC4 exposure into two
equal exposures.  As a result, the 8.0\,\micron\ observations
consist of four pairs of 8 dithers of 46.8 sec exposures. 

The results of the \emph{Spitzer} observations of the entire
population of IC\,348 members is presented in \citet{lada06}.  For
consistency, we processed the Basic Calibrated Data from the GTO
program and produced point source catalogs using the c2d pipeline. The
c2d pipeline uses the the c2d mosaicking/source extraction software,
c2dphot \citep{harvey04}, which is based on the mosaicking program
APEX developed by the \emph{Spitzer} Science Center and the source
extractor Dophot \citep{shechter93}.  See \citet{evans06} for a
detailed description of the c2d data products. We searched the c2d and
GTO IRAC point source catalogs and found fluxes for 127 of the 143
objects listed in Table~\ref{xgfp_parameters} (i.e., 16 of the
periodic objects fall outside the c2d/GTO IRAC maps of Perseus).

\subsection{Complementary data}

In addition to the rotation periods and the IRAC photometry, we have
collected the following complementary data for our stars in
Table~\ref{xgfp_parameters}. First, with the same telescope used to
obtain the time series photometry, we obtained RI absolute photometry
for our field. We performed PSF fitting photometry with the standard
IRAF implementation of the DAOPHOT \citep{stetson87} and used Landolt
standards for calibration \citep{landolt92}.  We report RI data for
all but a few stars listed in Table 1 that fall outside the field due
to a small pointing error.  Also, we have collected the 2MASS J, H and
K magnitudes for all but 3 very faint stars. Finally, we have collected 
all the spectral types available for our sample, covering 92 objects 
from \citet{luhman03}.  All the complementary data are also listed in
Table~\ref{xgfp_parameters}.

\section{Testing the Disk Regulation Paradigm}\label{rotation}

\subsection{Identification and Classification of Mid-IR Excess}\label{diskid}

In recent \emph{Spitzer} studies of circumstellar disks, different
groups adopt different disk identification criteria.  An effective and
reliable method of disk identification is crucial to this type of
survey, especially when dealing with a small sample size, such as is
the case with the IC\,348 rotation periods.  A near-100\% recovery of
disks with few or no false positives using IRAC photometry is
possible. Here we discuss the disk-identification criteria used in two
studies particularly relevant to this paper: The study of the
correlation between stellar rotation and IRAC excess in Orion
\citep{rebull06} and the \emph{Spitzer} observations of all the
confirmed members of IC\,348 \citep{lada06}. Bare stellar photospheres
have an IRAC color $\sim$\,0.0, while stars with an IR excess have
positive colors. The broader the color baseline used, the larger the
mean IR excess of the stars with disks; therefore, the IRAC color that
provides the clearest separation between stars with and without disks
is [3.6]--[8.0]. For their study of periodic stars in the ONC,
\citet{rebull06} adopt the color [3.6]--[8.0] $>$ 1.0 as the criterion
for disk identification. This boundary is chosen based on the shape of
the [3.6]--[8.0] histogram of their sample, which shows a clear
deficit of periodic stars around this color. \citet{lada06} use a
criterion based on the slope, $\alpha$, of a power law fit to the four
IRAC bands.  From the comparison to disk models, they identify objects
with $\alpha$ $>$ --1.8 as optically thick disks.  Based on the
predicted slope of an M0 star and the typical uncertainty in the power
law fit, they identify objects with $\alpha$ $<$ --2.56 as bare
stellar photospheres.  Objects with intermediate slopes,
--2.56\,$>$\,$\alpha$\,$>$\,--1.8, are termed ``anemic disks'' and are
interpreted as optically thin disks or disks with inner holes.

\begin{figure}
%\figurenum{5,6}
\epsscale{1}
\plottwo{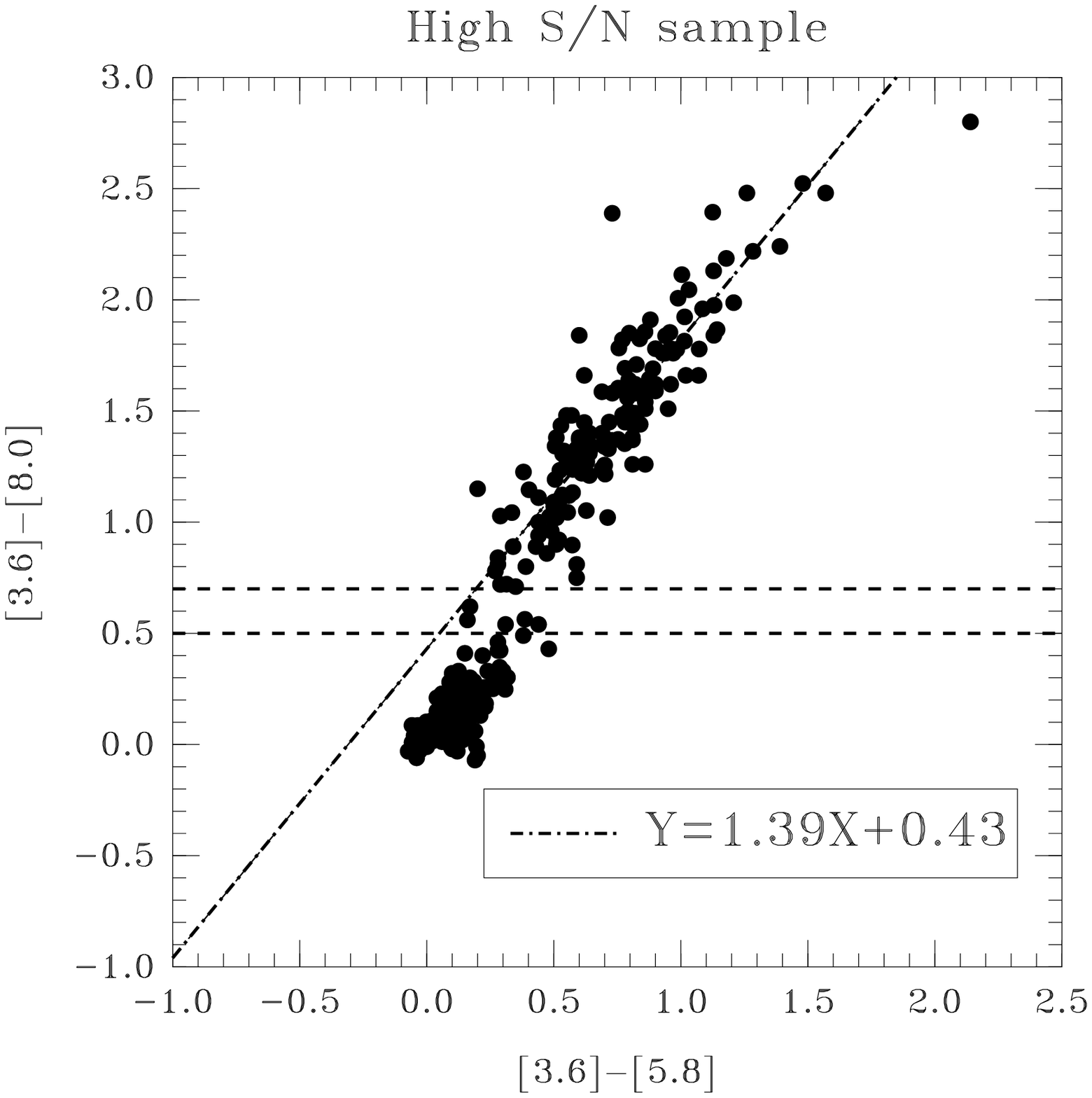}{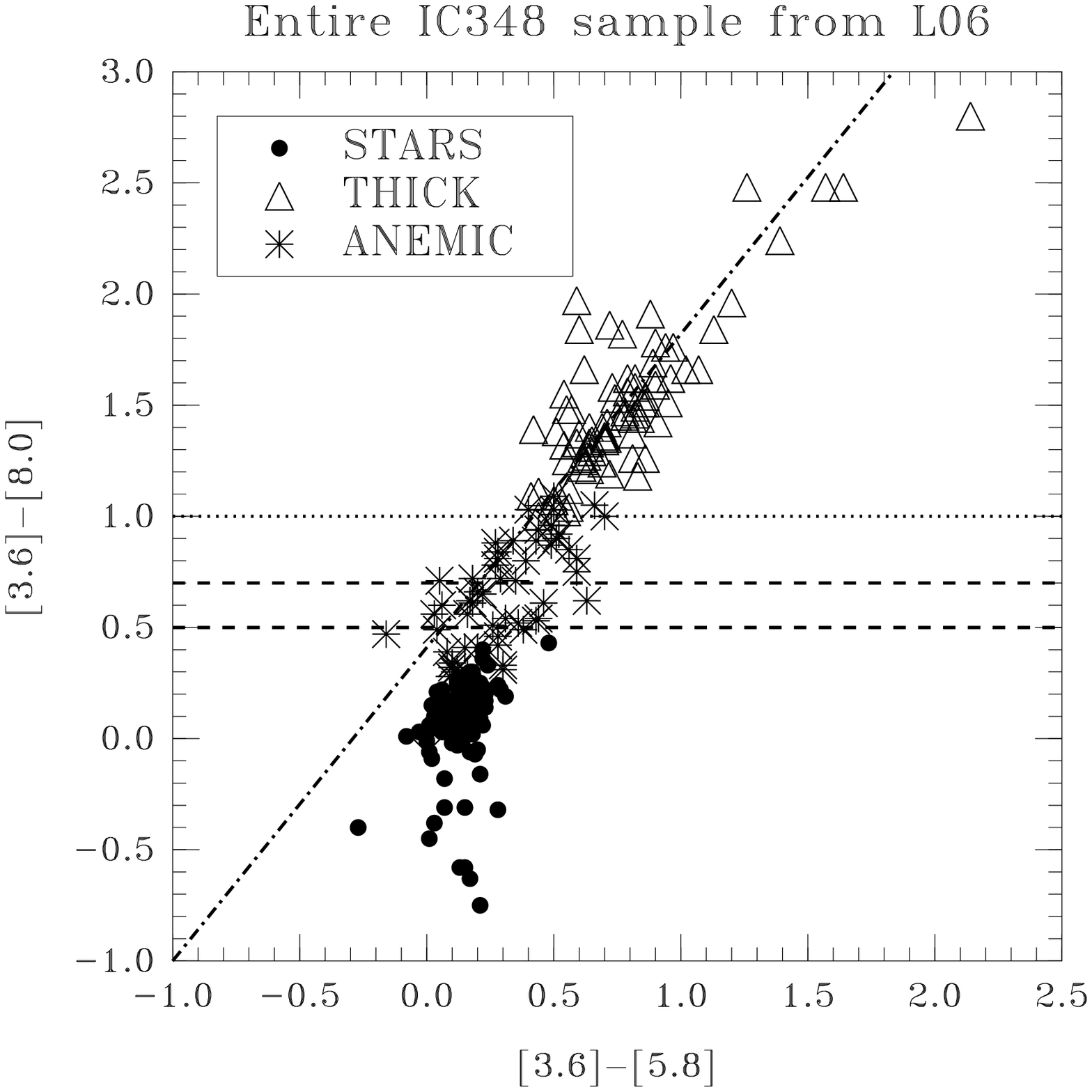}
\caption{The plot on the left shows the [3.6]--[8.0] {\it
vs}. [3.6]--[5.8] colors of 435 PMS stars collected from the
literature with photometric errors less than 0.1\,mag. Objects with
[3.6]--[8.0] $>$ 0.7 are stars with significant IR excess indicating
the presence of a disk. Objects with [3.6]--[8.0] $<$ 0.5 are
consistent with stellar photospheres. Only $\sim$\,1\% of the stars
have 0.7 $>$ [3.6]--[8.0] $>$ 0.5. The dash-dotted line is a linear
fit to the stars with disks. The plot on the right shows the entire
sample of IC\,348 members studied by \citet[][L06 in this
figure]{lada06}. Some of the objects classified as ``anemic disks'' by
L06 seem to be photospheres of late type stars.  The
dotted line correspond to the disk identification criteria adopted by
\citet{rebull06}.\label{colorcolordisk}}
\end{figure}

We find that the disk criteria adopted by the Lada and Rebull groups
identify a slightly different set of stars. In this section, we
propose alternative disk identification criteria and explore the
differences between our criteria and those adopted by the other
groups. In order to construct our disk-identification criteria, we
first collect IRAC colors for a large sample of PMS stars with high
signal to noise ratios (S/N) from the literature. We construct our
sample from the classical and weak-lined T\,Tauri stars (cTTs and
wTTs) studied by \citet{hartmann05}, \citet{lada06},
\citet{padgett06}, and Cieza et al. (2006, in prep). To minimize the
uncertainties, we restrict the sample to the 435 objects for which
reported photometric errors are less than 0.1\,mag and explore the
location of this sample in different color-color diagrams.  We find
that the [3.6]--[8.0] {\it vs}. [3.6]--[5.8] diagram shown in the left
panel of Figure~\ref{colorcolordisk} provides the best separation of
stars with an without a disk.  This combination of colors results in a
well-defined locus of stellar photospheres and of cTTs wTTs disks. We
identify the stars with [3.6]--[8.0] $>$ 0.7 located along the
dash-dotted line as stars with disks and stars with [3.6]--[5.8] $<$
0.5 and [3.6]--[8.0] $<$ 0.5 as diskless stellar photospheres. Only
$\sim$1$\%$ of the sample (5/435) have 0.7 $>$ [3.6]--[8.0] $>$ 0.5,
where the colors of the weakest disks seem to overlap with the colors
of the redder stellar photospheres. We note that the few photospheres
slightly redder than the main clump correspond, for the most part, to
M2--M6 stars.

Incidentally, a linear fit to the sample of stars with disks yields
\begin{displaymath}
([3.6]-[8.0])_{{\rm DISKS}} = 1.39\pm0.05 \times
  ([3.6]-[5.8])_{{\rm DISKS}} + 0.43\pm0.03
\end{displaymath}
This relationship extends over one magnitude in the [3.6]--[5.8] color
and two magnitudes in the [3.6]--[8.0] color with a 1\,$\sigma$
dispersion of 0.15\,mag. The tightness of this \emph{IRAC locus of T\,Tauri star
disks} is comparable to that of the loci of cTTs in the near-IR
defined by \citet{meyer97}. Therefore, one can use this convenient IRAC
locus of disks as an initial tool to identify wTTs and cTTs candidates
in any IRAC field.  One might expect contamination from galaxies with
IRAC colors similar to stars in this locus.  However, we found that 
in the fields observed by the Spitzer Wide-area InfraRed Extragalactic 
survey (SWIRE), only about $\sim$15\% of the objects fell within a 
3$\sigma$ distance from this IRAC locus.

In order to compare our disk-identification criteria to those adopted
by \citet{lada06} and \citet{rebull06}, we plot the [3.6]--[8.0] {\it
vs}. [3.6]--[5.8] colors reported by the Lada group for their entire
sample of IC\,348 objects (288 stars) with different symbols,
indicating the classification given in their study
(Figure~\ref{colorcolordisk}, right panel). This figure shows that the
disk-identification criterion adopted by Rebull and collaborators
(i.e. [3.6]--[8.0] $>$ 1.0) selects a sample of stars which has an
almost one-to-one correspondence with the objects classified as
optically thick disks by the Lada group, meaning the Rebull criterion
misses disks with weaker mid-IR signatures. On the other hand, Lada
and collaborators classify as anemic disks both objects that belong to
our IRAC locus of T\,Tauri star disks and some objects which have
colors consistent with bare stellar photospheres (i.e., the Lada
criteria overestimate the number of anemic disks). We believe this is
a consequence of adopting criteria for disk identification which are
independent of spectral type. In fact, most of the objects classified
as anemic disks by \citet{lada06} which do not satisfy our disk
identification criteria are late M stars. In addition, the scatter in
the right panel of Figure~\ref{colorcolordisk} is dominated by the
photometric error in 8.0\,\micron\ fluxes of the faintest stars, which
also comprise the handful of objects with nonphysical negative
[3.6]--[8.0] colors.

\begin{figure}
%\figurenum{7}
\epsscale{1}
\plotone{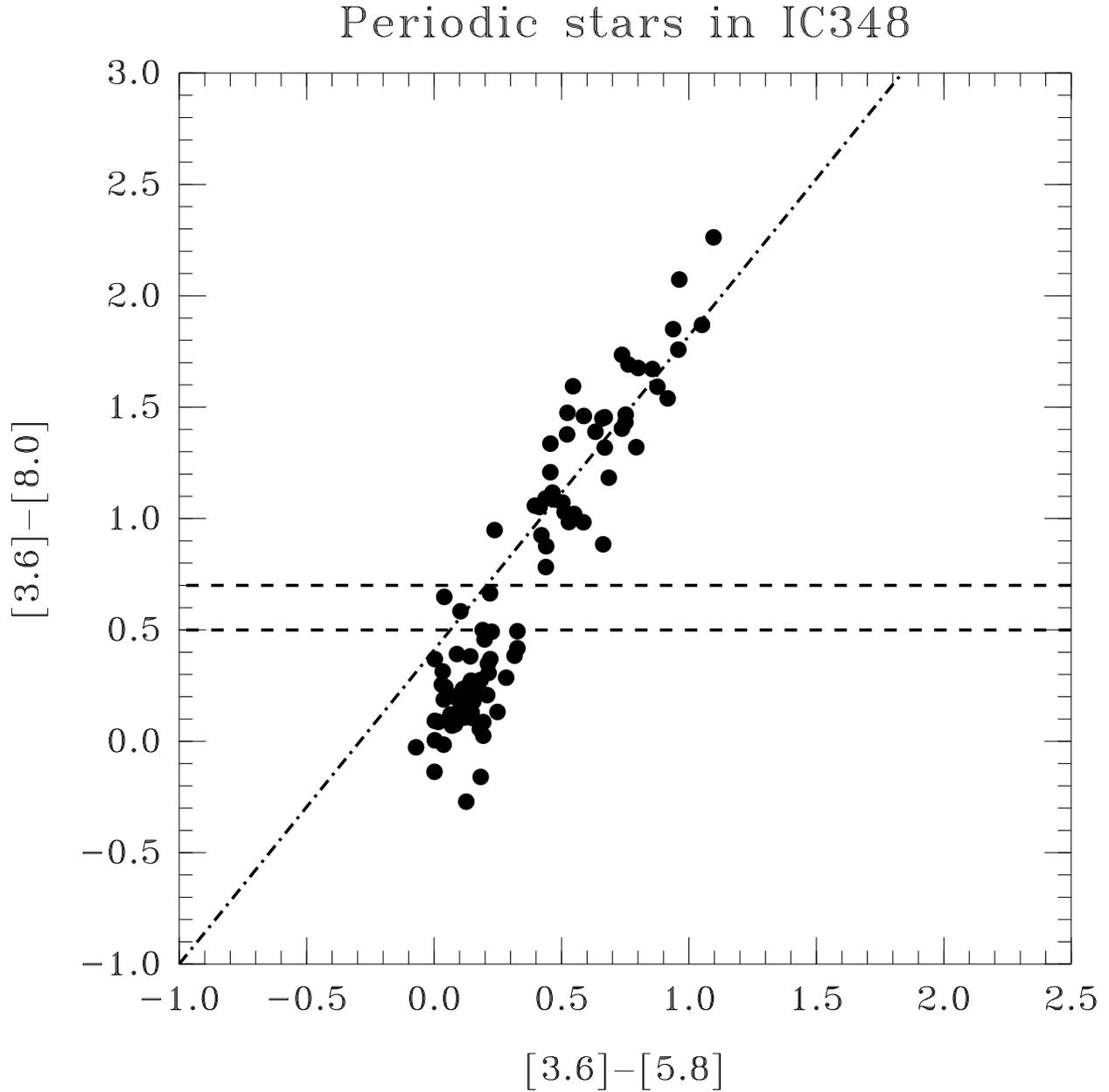}
\caption{The [3.6]--[8.0] {\it vs}. [3.6]--[5.8] colors of the periodic
stars in IC\,348.  Stars with [3.6]--[3.8] colors $>$ 0.7 possess disks
while stars with [3.6]--[3.8] colors $<$ 0.5 are diskless. Only three
objects show a somewhat ambiguous disk identification.\label{ic348diskcolor}}
\end{figure}

Figure~\ref{ic348diskcolor} shows the [3.6]--[8.0] {\it
vs}. [3.6]--[5.8] colors of our periodic sample in IC\,348, restricted
to the 109 stars in Table~\ref{xgfp_parameters} with 3.6, 5.8, and
8.0\,\micron\ fluxes and periods $<$ 15 days. We find that 40 stars
have a color [3.6]--[8.0] $>$ 0.7, indicating the presence of a disk,
and that only 3 stars have borderline colors, 0.7 $>$ [3.6]--[8.0] $>$
0.5.  Two of the objects with intermediate colors (ID= 72 and 111) are
faint M4.75--M5 stars; therefore, we classify them as diskless
stars. The other object (ID=88) is of unknown spectral type. We
searched the catalogs from the c2d Legacy Project and found that this
object has a 24\,\micron\ flux of 23.1\,mJy, which corresponds to a
24\,\micron\ excess of over 4 magnitudes. Thus, we classify this
object as possessing a disk, leading to a total number of disks in our
sample of 41 and a disk fraction of 38$\%$\,$\pm5\%$ (41/109).

Given the fact that only 3 objects in our sample have a somewhat
ambiguous disk identification, we estimate our disk census to be both
complete and reliable at the $\sim$97$\%$ level.  In contrast,
\citet{hillenbrand98} show that color-color diagrams combining optical
and near-IR data can only identify $<$\,70\% of the accretion disks in
a given sample and that this efficiency decreases for very low-mass
stars (e.g. spectral types later than M2). Also, \emph{Spitzer} can
identify stars with low accretion rates and disks with inner holes
which have no near-IR excess \citep[e.g.][]{padgett06}, making
\emph{Spitzer} data (combined with effective disk-identification
criteria, as described above) a more effective tool for disk
identification than any ground-based near-IR observations.

\begin{figure}
%\figurenum{8}
\epsscale{1}
\plotone{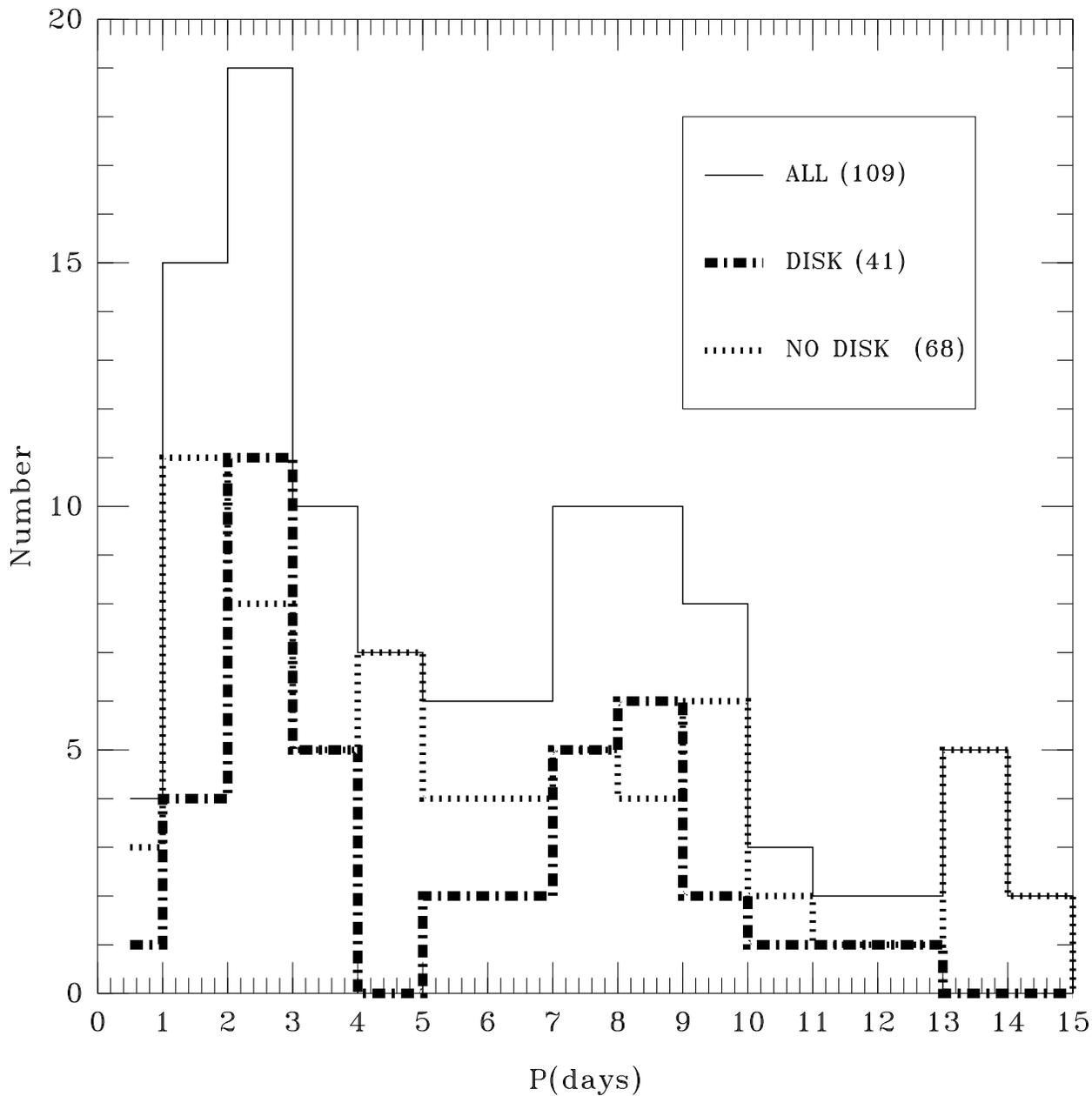}
\caption{Period histogram for 109 IC\,348 stars with rotation
  periods $<$ 15 days and  [3.6]--[8.0] data.  The three
  different lines represent all stars (solid), stars with an IR excess
  indicating the presence of a disk (dot-dash line), and stars with no
  detected disk signature (dotted line).  A very clear bimodal
  distribution is seen for stars with disks; there is no  clear 
  correlation between the presence of a disk and the rotation period for 
  the stars in IC\,348.\label{periodhist}}
\end{figure}

\begin{figure}
%\figurenum{9}
\plotone{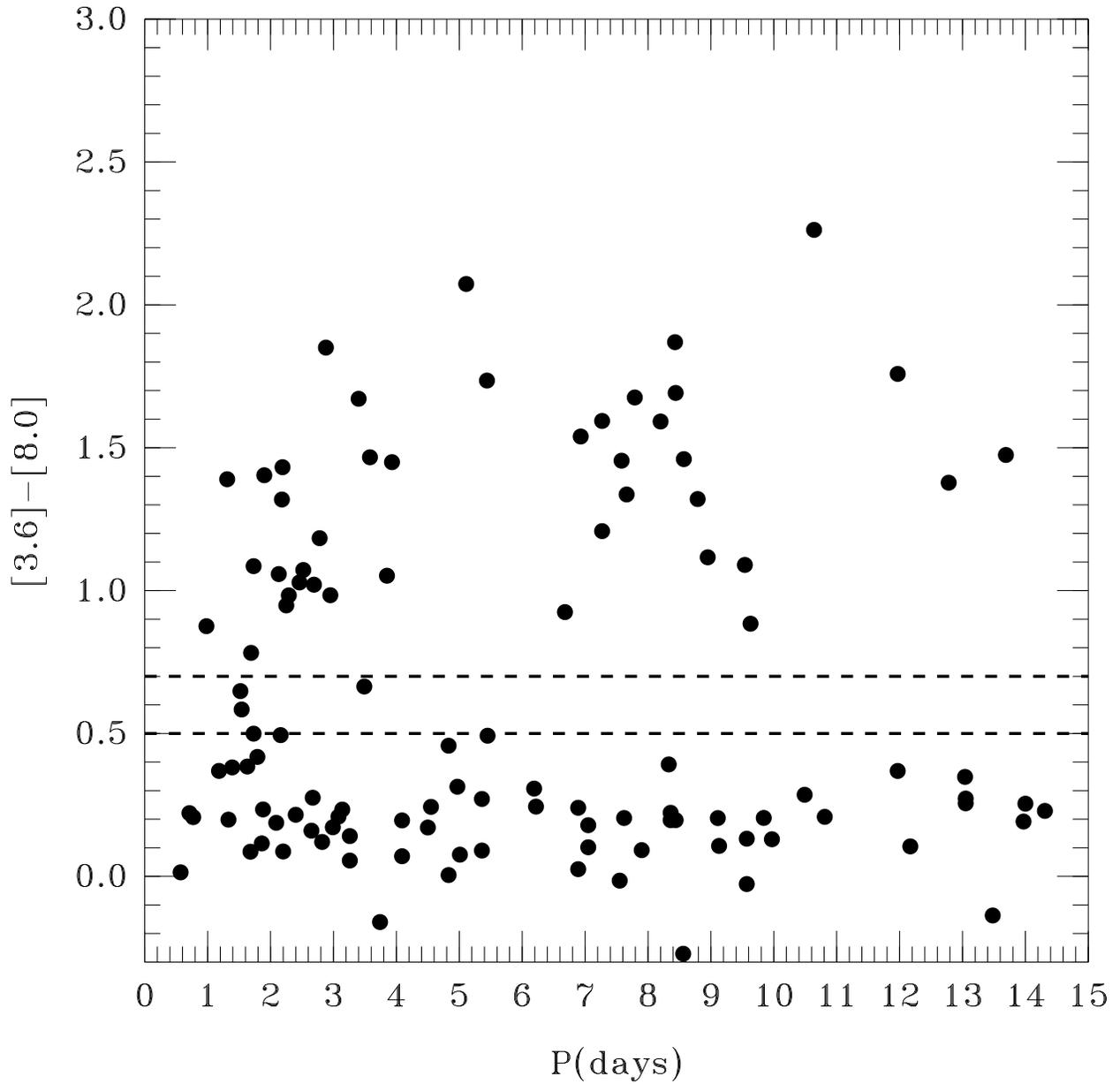}
\caption{A plot of [3.6]--[8.0] color vs. period. We find no evidence
for a correlation between period and the presence of an IR excess or
the magnitude of the excess. A standard Spearman test yields over a
84$\%$ chance that the quantaties are completely uncorrelated.\label{colortest}}
\end{figure}

Having accurately determined which stars have disks, one can then
conduct a comprehensive search for any correlation between stellar
rotation period and the presence of a disk.  The period distributions
for stars with and without a disk are shown in
Figure~\ref{periodhist}, while Figure~\ref{colortest} shows the
[3.6]--[8.0] color as a function of rotation period.  The most
striking feature of Figure~\ref{periodhist} is that stars with disks
show a bimodal distribution even more-clearly defined than that seen
in the entire sample.  This fact goes directly against the first order
prediction of the current disk-locking paradigm in which the bimodal
distribution itself is a manifestation of two populations of stars,
one with disks (slow rotators) and another without disks (fast
rotators). Similarly, Figure~\ref{colortest} shows no evidence of any
correlation between period and the \emph{magnitude} of the IR
excess. A standard Spearman test yields over an 84$\%$ chance that the
quantities are completely uncorrelated. However, before drawing any
further conclusions from Figures ~\ref{periodhist} and
\ref{colortest}, we would like to explore and attempt to disentangle
another variable which has been claimed to correlate with rotation
period: Stellar mass.
 
\subsection{Mass Dependence}

Following H00, most authors have divided their rotation period
distributions into ``high mass'' and ``low mass'' stars. The H00
masses come from \citet{hillenbrand97} and were obtained using the
evolutionary tracks by D98.  The 0.25\,$M_{\odot}$ division should be
regarded as a nominal value throughout the paper since PMS stellar
masses are highly model dependent.  We point out that D98 models yield
systematically lower masses than other widely used evolutionary models
such as those presented by \citet{baraffe98} and
\citet{siess00}. According to the D98 models, for ages less than
several million years, the 0.25\,$M_{\odot}$ division corresponds to
the M2 spectral type.  Therefore, other authors
\citep[e.g][]{rebull01} have used spectral type as a proxy for stellar
mass. Optical colors (${\rm R_{C}}$--${\rm I_{C}}$) have also been
used to differentiate between low- and high-mass stars
\citep[e.g.][]{lamm05}.

Ever since a statistically significant number of rotation periods
became available, it has been suggested that rotation periods are highly
dependent on mass. H00 argued that stars more massive than
$\sim$0.25\,$M_{\odot}$ show a bimodal distribution, while lower-mass
stars show a more uniform distribution and a lack of fast rotators
(P\,$<$\,2 days).   H00 attributed the lack of low-mass fast
rotators to a deuterium-burning phase that temporarily halts the
contraction of low mass stars. However, deeper observations of the ONC
(e.g. H02) later showed a large population of fast-rotating, low-mass stars. 
In fact, H02 argue that low-mass stars rotate significantly faster than
high-mass stars and that the period distribution of low-mass stars
peaks at $\sim$2 days (i.e. the apparent lack of low-mass fast rotators
previously reported was a selection effect). Using ${\rm
R_{C}}$--${\rm I_{C}}$ color criteria, \citet{lamm05} also find that
low-mass (${\rm R_{C}}$--${\rm I_{C}} > 1.3$) stars rotate, on
average, faster than ``high'' mass ( ${\rm R_{C}}$--${\rm I_{C}} < 1.3$)
stars. However, other authors have cast doubts on this
conclusion. \citet{rebull01} finds that the periods of high-mass stars
(spectral type earlier than M3) and low-mass stars (spectral types M3
and later) in the Orion Flanking Fields are statistically consistent
with each other. However, as mentioned earlier, \citet{herbst05}
regard the \citet{rebull01} sample as too heterogeneous in terms of
age for any period dependence on disk properties or mass to be
observable.

Naturally, we would like to explore the mass dependence in our period
distribution.  In order to divide our sample into ``high'' and ``low''
mass stars, we take advantage of the large amount of information
collected in Table~\ref{xgfp_parameters}. Most of the stars in our
sample (92/143, 64\%) have spectral types from \citet{luhman03}. We
classify stars with spectral type M2 and later as low-mass stars and
stars with spectral types earlier than M2 as high-mass stars.  For
stars without spectral types, we adopt the following procedure. First,
we classified the objects based on their [3.6]--[4.5] {\it vs}.
[3.6]--[8.0] colors.  Sources with [3.6]--[4.5]~$<$~0.2 and
[3.6]--[8.0]~$<$~0.5 are consistent with bare stellar
photospheres. For these objects, we obtain photometric approximations
of spectral types by fitting the RIJHK and IRAC magnitudes to stellar
models of different spectral types corrected for extinction.

The stellar models were constructed as in \citet{cieza05}, from
optical and near-IR broadband colors from \citet{kenyon95} tied to
IRAC colors based on Kurucz models \citep{kurucz93}. We adopt the
extinction corrections also listed in \citet{cieza05}. For stars with
[3.6]--[4.5]~$<$~0.2 and [3.6]--[8.0]~$>$~0.5 (e.g. stars without
4.5\,\micron\ excess but with significant 8.0\,\micron\ excess), we
follow the procedure describe above, but without fitting the
8.0\,\micron\ or 5.6\,\micron\ data points (in practice, the weights
of the IRAC data points in the fit are set by the corresponding
``excess''). We tested our procedure with 58 stars with spectral types
and [3.6]--[4.5]~$<$~0.2 and find that, remarkably, our photometric
spectral types typically agree with those reported by \citet{luhman03}
to within a spectral subtype. Namely, {\bf MEAN}$({\rm
Spt_{Luhman}-Spt_{phot}}$)\,=\,0.26 and {\bf MEAN}${\rm
(ABS(Spt_{Luhman}-Spt_{phot}}))$=0.69, where one spectral type subclass
equals unity and spectral types are ordered from early to late.

\begin{figure}
%\figurenum{10,11}
\epsscale{1}
\plottwo{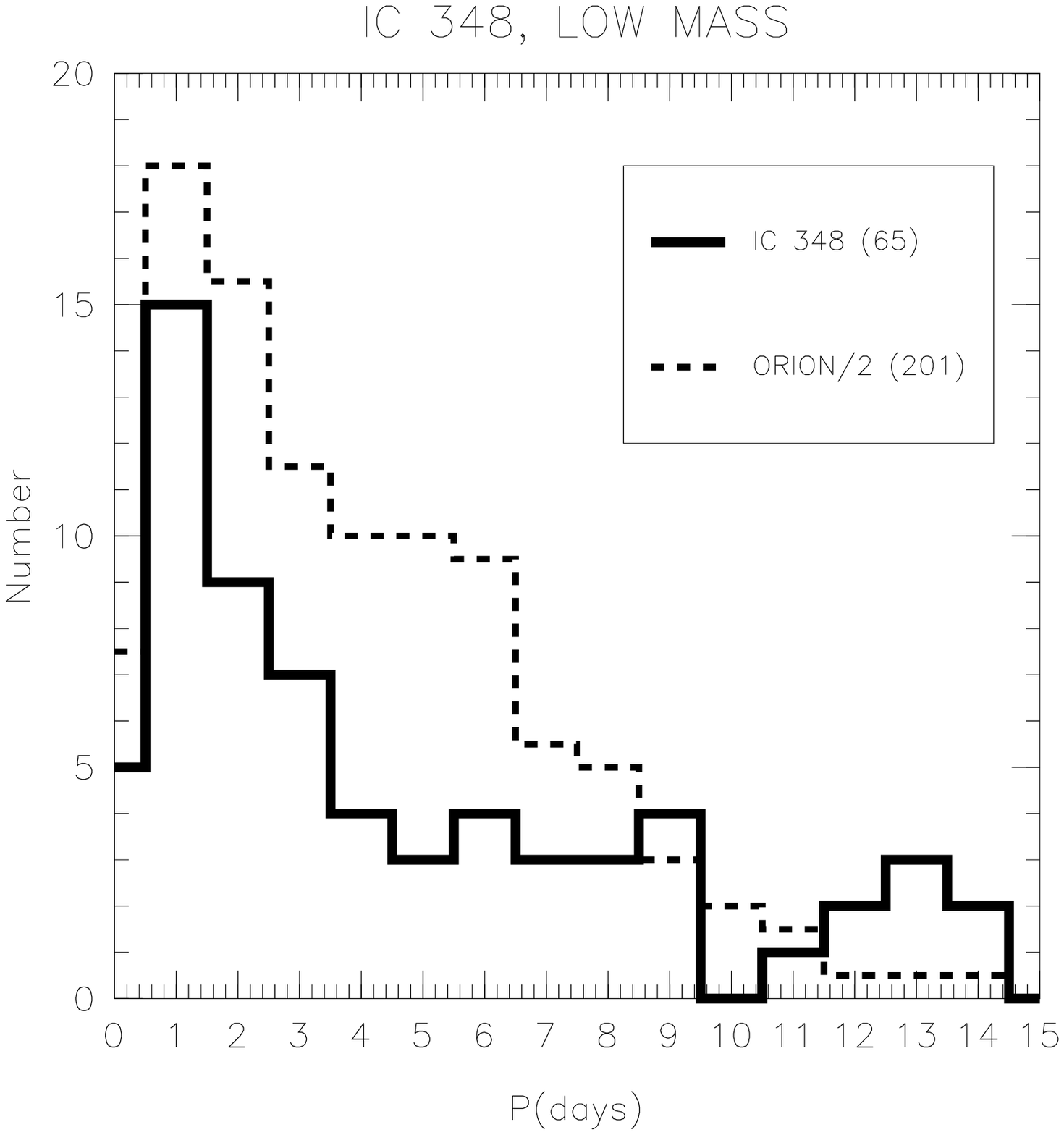}{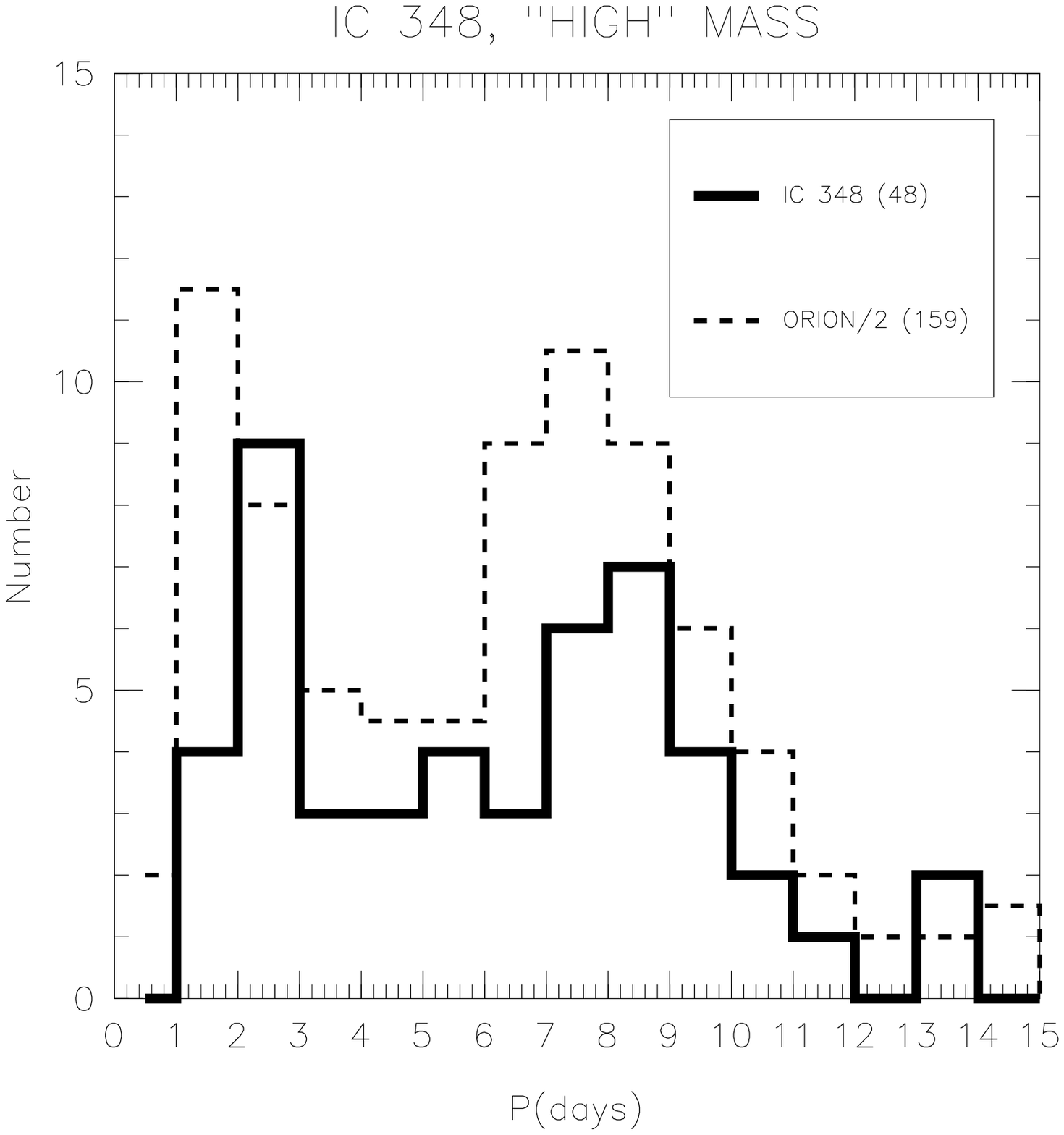}
\caption{Period histograms for low and high mass stars in the IC\,348
cluster with \emph{Spitzer} data.  The period distributions resemble those seen in the 
heart of the Orion Nebula Cluster (ONC) by H02, which are shown for comparison
scaled down by a factor of two. Stars  estimated to be less massive than
0.25\,$M_{\odot}$ show a unimodal distribution dominated by fast
rotators (P\,$\sim$\,1--2 days), while stars estimated to be more
massive than 0.25\,$M_{\odot}$ show a bimodal distribution with peaks
at $\sim$2 and $\sim$8 days.\label{highlowhists}}
\end{figure}

Finally, objects with [3.6]--[4.5]~$>$~0.2 and [3.6]--[8.0]~$>$~0.5
are likely to be cTTs with thick inner disks
\citep[e.g.][]{lada06}. \citet{meyer97} show that cTTs occupy a
very-well-defined locus in the \emph{dereddened} H--K {\it vs}. J--H
diagram. We take advantage of this fact and estimate the ${\rm A_{V}}$
of the few objects with [3.6]--[4.5]~$<$~0.2 and unknown spectral types
by dereddening them to the the locus defined by \citet{meyer97}.
Using the ${\rm A_{V}}$'s estimated in this way, we deredden their
${\rm R_{C}-I_{C}}$ colors.  We classified stars with ${\rm
(R_{C}-I_{C})_{o}}$\,$<$\,1.3 as high-mass stars and stars with ${\rm
(R_{C}-I_{C})_{o}}$\,$>$1.3 as low-mass stars (where a ${\rm
R_{C}-I_{C}}$ color of 1.3 corresponds to a spectral type of M2). The
period distributions for low- and high-mass stars are shown in
Figures~\ref{highlowhists}. We find that the period distributions are
remarkably similar to those presented by H02 for the heart of the
ONC. Stars estimated to be less massive than 0.25\,$M_{\odot}$ show a
unimodal distribution dominated by fast rotators (P\,$\sim$\,1--2 days)
and a tail of slow rotators, while stars estimated to be more massive
than 0.25\,$M_{\odot}$ show a bimodal distribution with peaks at
$\sim$2 and $\sim$8 days.  Thus, our results confirm the strong
dependence of stellar rotation on mass, as observed by H02 in the heart
of the ONC and by \citet{lamm05} in NGC\,2264.

As mentioned in section 2, \citet{lamm05} find that the peaks in the
period distribution observed in NGC\,2264 are shifted toward shorter
periods with respect to those seen the ONC. These shifts are
interpreted as evidence of angular momentum evolution between the the
ONC (age $\sim$1\,Myrs) and NGC\,2264 (age $\sim$\,2--4\,Myrs).  Do we
see the same angular momentum evolution between the ONC and IC\,348
(age $\sim$3\,Myrs)? On one hand, the location of the peaks in the
period distribution of the high-mass stars in IC\,348 suggests that
they have spun down with respect to those in the ONC. On the other
hand, the median period of the the low-mass stars in IC\,348 seems to
suggest that these stars have spun up with respect to to the low-mass
stars in the ONC. However, given the size of the sample and the
magnitude of the changes, none of the differences are statistically
significant.
  
\subsection{IR Excess--Rotation Period Correlation?}\label{ircorrelation}

\begin{figure}
%\figurenum{12}
\epsscale{1}
\plotone{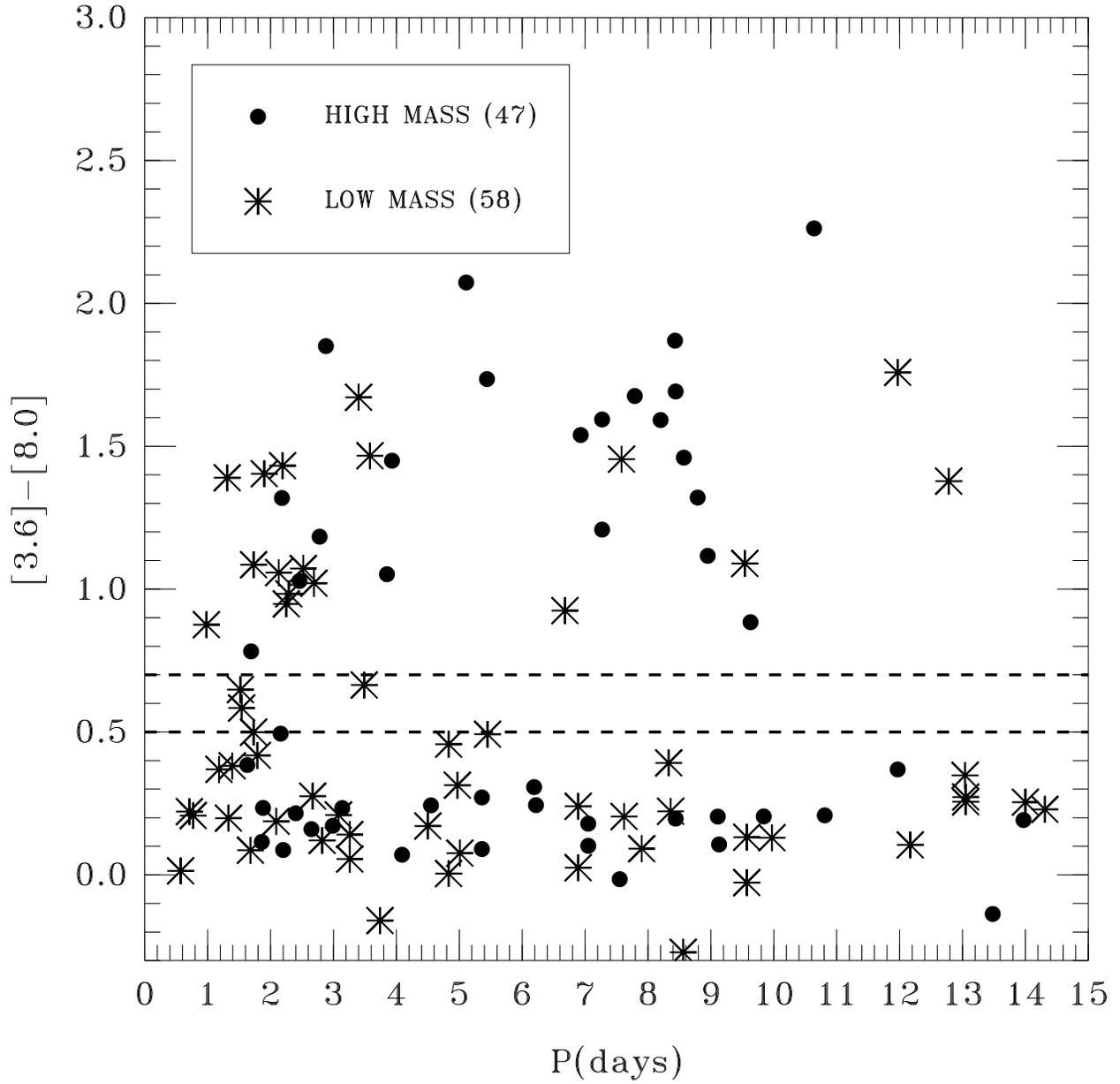}
\caption{[3.6]--[8.0] color {\it vs}. period for low- and high-mass
stars.  There is no significant correlation between IR excess and the
period of either type of star.\label{periodcolor}}
\end{figure}

Taking the low- and high-mass stars separately, is there any evidence
that stars with disks rotate slower, as a group, than stars without
disks? Given the substantial improvement in the disk identification
efficiency using IRAC observations over near-IR indicators, the
disk--period correlation reported by H00 and H02, if real, should be
readily-detectable in our data. Figure~\ref{periodcolor} shows the
[3.6]--[8.0] color {\it vs}. period for both low- and high-mass
stars. A standard Spearman test shows that there is a 68$\%$ chance
that the period of high-mass stars is completely uncorrelated with IR
excess. This chance drops to 23$\%$ for low-mass stars, but the
significance level is not anywhere near the level found by H02 for the
correlation between period and (I--K) excess in Orion stars (H02 find
that there is a $10^{-10}$ chance that period and (I--K) excess are
uncorrelated quantities).  Our results support the
claim made by \citet{littlefair05} that the correlation between period
and (I--K) is a secondary manifestation of the correlation between
near-IR excess and mass. Most of the power in the weak correlation
between period and IRAC excess comes from the fact that the 5 fastest
rotators (all of them low-mass objects) show little or no excess. We
discuss this point in Section~\ref{fraction}.  

Low-mass stars show a unimodal distribution peaking at $\sim$1--2 days
and a tail of slow rotators. According to the disk-braking model, one
would expect stars without disks to be concentrated around the
1--2-day peak and stars with disks to be concentrated in the
long-period tail.  We find no evidence of that being the case.  If
anything, we find the opposite to be true. The median period of the
sample of low mass objects is 3.58 days. The disk fraction of the
objects rotating faster than the median is 38$\%$\,$\pm8\%$ (11/29),
while the disk fraction of the objects rotating slower than the median
is 21$\%$\,$\pm7\%$ (6/29). High-mass stars show a bimodal
distribution peaking at $\sim$2 days and $\sim$8 days. In the context
of the disk-braking model, one would expect, for the most part, stars
populating the short-period peak to be objects without disks and the
stars populating the long-period peak to be stars with disks.  Again,
we find no significant evidence for disk braking. We find that
39$\%$\,$\pm10\%$ (9/23) of the high-mass stars rotating faster than
the median (P\,=\,6.2 days) have disks, while a marginally higher
fraction of stars rotating slower than the median, 50$\%$\,$\pm10$
(12/24), have disks.

\section{Discussion}

The fact that we find no evidence for a correlation between period and
the presence of IR excess represents a serious challenge to the
current paradigm of the evolution of angular momentum of PMS stars,
which relies heavily on disk braking to explain the presence of slowly
rotating PMS and ZAMS stars.  It could be argued that the
8.0\,\micron\ excess is \emph{too good} a disk indicator, enabling
detection of disks with inner holes too large for the disk to still be
locked to the star. However, disk models show that stars with inner
holes have lower 8.0\,\micron\ excesses than stars with disks extending
inward to the dust sublimation temperature ( e.g. \citet{padgett06}
models of wTTs {\it versus} \citet{cieza05} models of cTTs).
Therefore, even if a diversity of inner hole sizes were responsible
for masking the disk--period correlation in the histograms in Figure
\ref{periodhist}, one would expect periods to correlate with the
\emph{magnitude} of the 8.0\,\micron\ excess (one would expect
stars with strong 8.0\,\micron\ excess to rotate slower than stars
with weak or no 8.0\,\micron\ excess). As illustrated by
Figure~\ref{colortest}, we find no evidence of such a correlation
either.  Throughout the paper, we use the 8.0\,\micron\ excess as a disk
indicator because, as discussed in Section~\ref{diskid}, it provides
the best separation between stars with and without disks; however, we
see no period--IR excess correlation using the excess at any of the
IRAC bands as a disk indicator.

There is currently no quantitative theory for disk braking; however,
some authors have tried to make some quantitative analysis based on
the derived distributions of angular momentum for stars of different
ages. \citet{herbst05} construct the distributions of the specific
angular momentum of the ONC, NGC\,2264, and a sample of main sequence
(MS) stars, combining objects from the Pleiades, $\alpha$\,Per, and
the IC\,2602 clusters. Specific angular momentum, \emph{j}, is
obviously a more relevant quantity than period to study the evolution
of angular momentum of PMS stars. However, its calculation requires
the knowledge of the stellar radius
(e.g. \emph{j}\,$\propto$\,$R^{2}$/P). To that end, \citet{herbst05}
assume a mean radius of 2.09\,${\rm R}_{\odot}$ for stars in the ONC
and of 1.7\,${\rm R}_{\odot}$ for stars in NGC\,2264. They argue that
the ONC, NGC\,2264, and MS clusters represent a clear sequence in the
evolution of angular momentum corresponding to 3 different ages,
nominally: 1, 2 and 50 Myrs.

\citet{herbst05} suggest that the broad distribution of angular
momentum observed in their sample of MS stars can be explained by
assuming that 40--50$\%$ of the PMS stars conserve angular momentum
when they contract while the remainder of the stars, \emph{which
should already be slow rotators by the ages of the ONC and
NGC\,2264}, must stay disk-locked for up to 5 Myrs in order to account
for the slow rotators in the ZAMS sample (this is consistent with the
findings of \citep{rebull04}).  Were this the case, we should see
strong evidence for disk locking in our sample of PMS stars with
rotation periods and IR-excess measurements.  Since we see no such
evidence, our results are inconsistent with the \citet{herbst05}
claim, unless, for some reason, our sample is severely biased against
disk-locked stars.  We investigate the possibility of such a bias in
the next section.

\subsection{Sample bias?}\label{sampbias}

A critical point regarding the validity of the conclusions of this
paper is to what degree our sample of stars with rotation periods is
representative of the entire population of IC\,348 members.
Fortunately, the stellar population of IC\,348 and its disk properties
are very well characterized. As discussed in Section~\ref{oursample},
\citet{luhman03} present a detailed spectroscopic census of IC\,348
cluster members, complete for spectral type M8 and earlier, and
\citet{lada06} have recently presented \emph{Spitzer} observations of
all the 288 objects identified by \citet{luhman03} as IC\,348
members. Because the Lada results are based on the the same IRAC GTO
observations we use for disk identification, they provide a perfect
test to investigate the possibility of any strong disk fraction bias
in our sample of IC\,348 members with rotation periods with respect to
the entire cluster population.

\citet{cohen04} argue that, since none of the 28
periods they report correspond to known cTTs, rotation period samples
are heavily biased against accreting objects, where disk braking is
more likely to occur. In contrast, \citet{littlefair05} report that
43\% of their periods correspond to cTTs, a result which is not
significantly different from the overall cTTs fraction among IC\,348
members ($\sim40\%$). They argue that the drastic improvement in the
efficiency for obtaining rotation periods of cTTs in their work with
respect to the \citet{cohen04} results is due to the high sampling
density of their observations; cTTs are highly
non-periodic on time scales larger than a few days or
weeks. \citet{cohen04} periods come from an observing campaign
spanning several years with a very low temporal density, while
\citet{littlefair05} observations come from a 26-night campaign with a
typical sampling rate of one frame every 10 minutes.  Our observations
also had very high temporal density and yield a cTTs fraction of
$\sim$40$\%$.

Using the same disk identification criteria ([3.6]--[8.0] $>$ 0.7) for
the entire sample of IC\,348 members studied by \citet{lada06} and for
the periodic sample listed in Table~\ref{xgfp_parameters}, we find the
disk fractions to be the same within statistical errors,
41$\%\pm2.7\%$ vs. 38$\%\pm5\%$, respectively. Since there is no
evidence that our sample is biased against stars with disks, we
conclude that our sample of stars with rotation periods is
representative of the entire population of IC\,348 members in terms of
their disk properties.  \citet{rebull06} reach the same
conclusion for the stars in Orion by noting that the [3.6]--[8.0] color
distribution of the periodic stars is statistically indistinguishable
from that of the entire population of Orion members. Therefore, we
argue that it is very unlikely that the lack of correlation between
rotation period and IR excess is due to a bias in the disk properties
of our sample.

\subsection{Previous Spitzer Results and Very Fast Rotators}\label{fraction}

Previous to this paper, the analysis of the periodic stars in Orion by
\citet{rebull06} is the only study that combines rotation periods and
\emph{Spitzer} observations.  They study a sample of 464 stars with
known rotation periods and [3.6]--[8.0] colors.  They note that the
cumulative distribution function of periods for stars with disks
(i.e., [3.6]--[8.0] $>$ 1) shows an inflection point around 1.8
days. Based on this inflection point, they divide their sample into
``short'' and ``long'' period stars.  They find that stars with
short periods are \emph{significantly} less likely to have a disk
than stars with long periods. However, they also find that for periods
$>$ 1.8 days, the period distributions for stars with and without
disks are statistically \emph{indistinguishable}.

Our results are consistent with those of the Rebull group. Although at
a lower significance level, due to the smaller size of our sample, our
results also suggest that there is a significant decrease in the disk
fraction at very short periods. As mentioned in
Section~\ref{ircorrelation}, most of the power in the weak correlation
between period and 8.0\,\micron\ excess for low-mass stars in IC\,348
arises from the fact that the 5 fastest rotators (P\,$<$\,1.2\,days)
have very little or no excess at all.  Using the same disk
identification criteria ([3.6]--[8.0] $>$ 1.0) and definition of
``short'' and ``long'' period stars adopted by \citet{rebull06}, we
find that only 12$\%$\,$\pm7\%$ (2/16) of the short-period (P$<$ 1.8
d) stars in IC\,348 have a disk, while 33$\%$\,$\pm5\%$ (31/93) of the
long-period (P $>$ 1.8 d) stars have a disk.  These disk fractions are
consistent with those seen in the ONC by Rebull and collaborators.

\begin{figure}
%\figurenum{13}
\epsscale{1}
\plotone{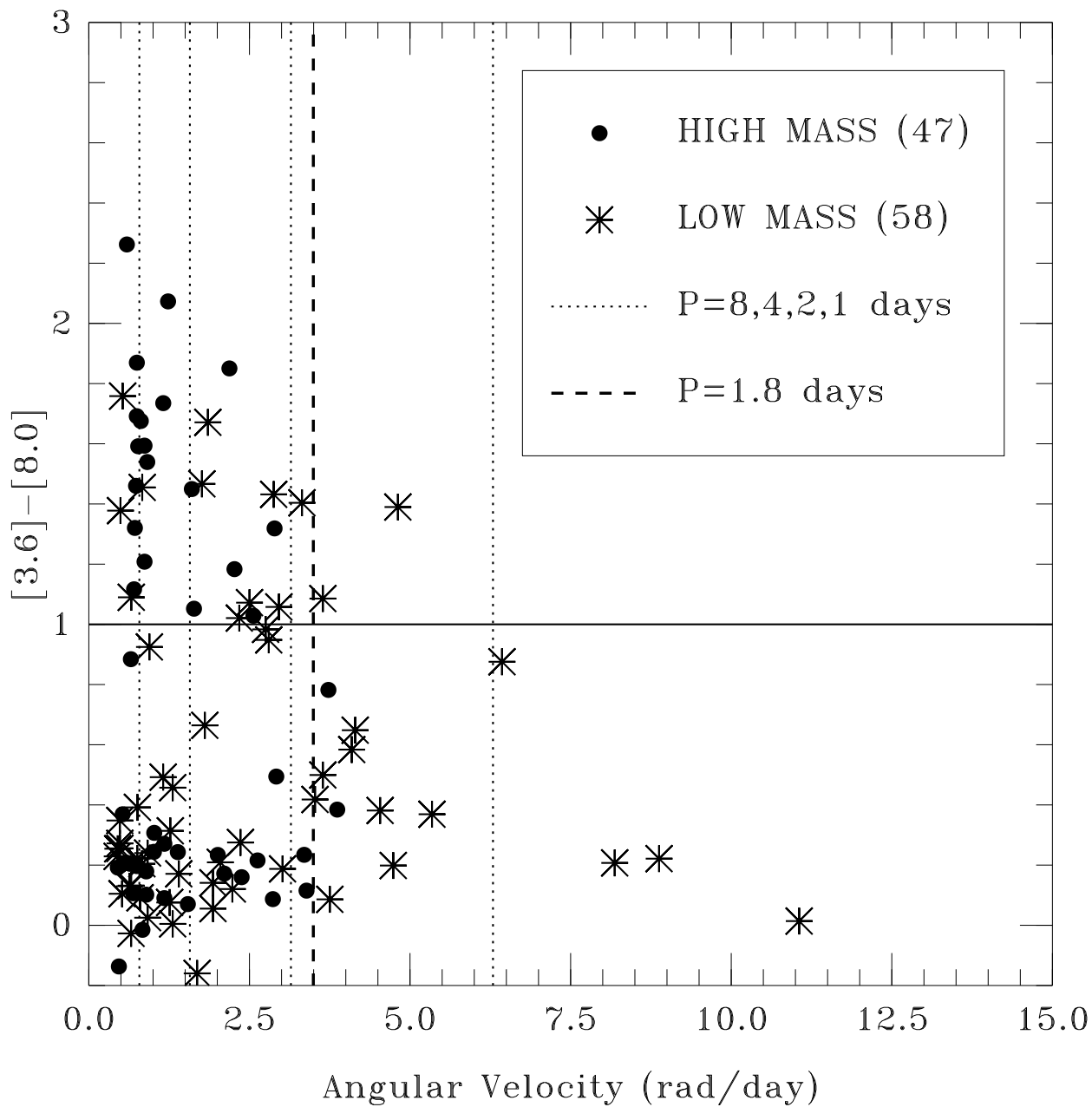}
\caption{[3.6]--[8.0] color {\it vs}. angular velocity for low- and
high-mass stars.  Stars with periods $\lesssim$ 1.5 days are
significantly less likely to have a disk than stars with longer
periods. The low disk frequency of very fast rotators is the only
feature of our sample that could \emph{potentially} be interpreted as
an evidence for disk braking, but a more rigorous analysis of this
result is necessary to determine its  significance.\label{veryfast}}
\end{figure}

The low disk fraction of fast rotators is evident in
Figure~\ref{veryfast}, where we plot angular velocity (2$\pi$${\rm P}^{-1}$)
as a function of [3.6]--[8.0] color in order to better resolve the
distribution of IR excesses at short periods. At first glance, the
upper envelope of the IR excess seems to correlate with angular
velocity. However, that is mostly due to the fact that most of the
objects are concentrated at low angular velocity.  The objects plotted
in Figure~\ref{veryfast} are the same objects plotted in
Figure~\ref{periodcolor}. From a statistical point of view, the
significance in the correlation between period and IR excess is
\emph{identical} to that of the correlation between angular velocity
and IR excess.  The fact that very fast rotators tend to have little
or no excess is also confirmed to a high level of significance by a
preliminary analysis of \emph{Spitzer} data on NGC\,2264, which will
be fully analyzed and presented in a follow-up paper (Cieza et al. 2007, 
Paper II).

The low disk frequency of very fast rotators is the only feature of
our sample that could \emph{potentially} be interpreted as an evidence
for disk braking. In the context of disk braking, stars that lost
their disks very early in their evolution are expected to become very
fast rotators.  The main challenge for disk braking seems to be the
large number of slow rotators without a disk. Qualitatively, it has
been proposed that the large number of slow rotators that show no
evidence of a disk are objects that have recently lost their disks and
have not had enough time to spin up considerably
\citep[e.g. H02,][]{rebull05, rebull06}.  This scenario would require
the spin-up timescale for the stars in the sample to be significantly
longer than the transition timescale from a massive inner disk capable
of disk braking to a disk tenuous enough to remain undetectable at
IRAC wavelengths.  However, we note that the population of fast
rotators (P $<$ 1.8 d) represents only a small fraction ($\sim$\,15\%)
of the entire sample of stars with rotation periods in both IC\,348
and the ONC.  In other words, the vast majority of stars with rotation
periods, about 85\% in each cluster, shows no correlation between
their rotation period and the presence of a disk.  This casts serious
doubts on the explanation discussed by the three groups mentioned
above that the large population of slow rotators with no mid-IR excess
is populated by stars which have just lost their disks.  This would
require $\sim$\,50\% of the entire sample of periodic stars in each
cluster to fall into this special regime.  A more quantitative test of
this scenario is underway and will be presented in Paper II.

Furthermore, we note that there are other possible explanations for
the low disk fraction of very fast rotators besides disk braking. For
instance, if the disk fraction of low mass stars, which tend to be
very fast rotators, is significantly lower than that of higher mass
stars, then an overall lower disk fraction of fast rotators is
expected.  Results from \citet{lada06} suggest that this might be the
case. They find that the fraction of \emph{optically thick} disks
decreases from 47\%\,$\pm$12\% for K6--M2 stars to 28\%\,$\pm$5\% for
M2--M6 stars. As discussed in Section~\ref{diskid} and illustrated by
Figure~\ref{colorcolordisk}, the criteria adopted by \citet{rebull06}
for disk identification selects a sample which has almost a one-to-one
correlation with the objects classified as optically thick disks by
the Lada group. Also, as noted by Rebull and collaborators, an
overabundance of close binaries among very fast rotators could account
for their low disk fraction.
                                                                                                            
\subsection{Are our results inconsistent with disk-braking?}\label{results}

How significant is the fact that we see no clear evidence of a
correlation between rotation period and IR excess (with the exception of
the fastest rotators as mentioned in the previous section)?  Is the lack of
evidence enough to exclude disk braking as a viable model?  At least
one line of argument suggests the contrary.  \citet{rebull04} perform
a series of Monte Carlo simulations of the evolution of rotation
periods for stars in the context of the disk-braking model. They
assume a Gaussian initial period distribution centered at 8 days with
a 1\,$\sigma$ dispersion of 4 days. In their models, they have a
population of disk-locked stars that contract at constant angular
velocity and a disk-free population which spins up as
P\,$\propto$\,$t^{-2/3}$ (appropriate for stars on convective
tracks). The only free variables in their model relate to the fraction
of disk-locked stars as a function of time.

After running a series of
simulations, they perform K--S tests to constrain the circumstances
under which a statistically significant correlation between period and
IR excess would be expected. They conclude, given a broad
distribution of initial periods, that the observational signatures of disk
braking are much less conspicuous than is usually assumed, \emph{even
with a perfect knowledge of which stars have a disk}. They find that,
unless the sample is very large ($>$\,500), the populations of stars
with and without disks only become statistically different if a
relatively large fraction ($\sim$30$\%$) of the stars is released
from their disks at an age $<$\,1\,Myr. This is because the effect of
disk locking is most important when the stars undergo rapid
contraction at very young ages and becomes much less important later
on as the contraction rate decreases.  Furthermore, the
\citet{rebull04} simulations can produce both a unimodal distribution
of periods similar to the one seen in low-mass stars in IC\,348 and a
bimodal distribution similar to that of the higher-mass stars.

These caveats prevent us from drawing any categoric conclusion
regarding the general validity of the disk-locking scenario. There is
currently no alternative model to disk braking to explain the
evolution of angular momentum of PMS stars, and there is some strong
evidence that magnetic star--disk interaction actually occurs in early
stages of the evolution of PMS stars (e.g. the presence of
highly collimated jets in cTTs and deeply embedded
objects). Therefore, we argue that a much more rigorous and
quantitative analysis than that presented herein is required before
the model can be regarded as inconsistent with the observational
data. The fundamental question now becomes whether disk braking, which
seems to be required to explain the angular momentum loss experienced
by a large fraction of PMS stars between the birthline and the ZAMS,
can occur early on in the evolution of PMS stars without leaving a
clear correlation between rotation period and IR excess at the ages of
clusters like IC\,348.

To address this question, a more rigorous model than the one used by
\citet{rebull04} could be created to investigate the range of
disk-braking parameters (disk-braking efficiency and the fraction of
regulated stars as a function of time) which would be allowed by the
observed period distribution of stars with and without disks in
clusters like IC\,348, NGC\,2264, and the ONC.  For instance, instead of
starting from an initial distribution of rotation periods, one could
use specific angular momentum, which is more physically meaningful and
removes most of the dependence on the initial masses of the objects
from the results.  The initial angular momentum of the cluster can be
constrained by the objects with a maximum current angular momentum
(since they will have evolved with the lowest angular momentum loss).
It would also be beneficial to incorporate into the models an age
spread in the stellar population and observational constraints on the
disk fraction as a function of age.  Finally, instead of assuming a
P\,$\propto$\,$t^{-2/3}$ relation for all unregulated stars, the
evolution of their periods can be calculated from the predicted
evolution of their radii by theoretical evolutionary tracks.

In Paper II, we plan to perform Monte Carlo simulations to determine
what initial conditions and disk-braking parameters might lead to the
current lack of a correlation between rotation period and IR excess in
PMS stars.  We will use models similar to \citet{rebull04} with the
improvements mentioned above to perform an analysis of the IC\,348
cluster, using data presented in Table~\ref{xgfp_parameters}, and
NGC\,2264 and the ONC, combining \emph{Spitzer} archival data and
periods published in the literature.

\section{Summary and Conclusions}

We have obtained time series photometry of the young stellar
cluster IC\,348 and measured 75 new rotation periods. Our results
increase the total number of known rotation periods in the cluster to
143.  We combined all published rotation periods in IC\,348 with
\emph{Spitzer} photometry (3.6, 4.5, 5.8, and 8.0\,\micron) in
order to test the disk braking paradigm, constructing a new, more
reliable set of criteria for disk identification with IRAC data.
 
We find that the IC\,348 rotation period distribution resembles that
seen in the heart of the ONC: Stars estimated to be less massive than
0.25\,$M_{\odot}$ show a unimodal distribution dominated by fast
rotators (P\,$\sim$\,1--2 days) and a tail of slow rotators, while
stars estimated to be more massive than 0.25\,$M_{\odot}$ show a
bimodal distribution with peaks at $\sim$2 and $\sim$8 days. We find
no evidence that the tail of slow rotators in low-mass stars or the
long-period peak of high-mass stars are preferentially populated by
stars with disks, a correlation which is predicted by the current
disk-braking paradigm. Also, we find no significant correlation
between period and the \emph{magnitude} of the IR excess, regardless
of the mass range considered.

Given the large improvement of IRAC observations over near-IR disk
indicators, our results support the claim made by \citet{littlefair05}
that the correlation between period and (I--K) excess reported by
several authors is a secondary manifestation of the correlation
between near-IR excess and mass.  We find that the disk properties of
our sample are indistinguishable from the disk properties of the
cluster as a whole and conclude that it is very unlikely that the lack
of a correlation between rotation period and IR excess is due to a
bias in the disk properties of our sample.  Finally, we find some
indication that the disk fraction might decrease significantly in
stars with very short periods (P\,$\lesssim$\,1.2 days). The fact that
very fast rotators tend to have little or no excess has already been
shown by \citet{rebull05} for stars in the ONC and has been confirmed
by a preliminary analysis of our ongoing work with objects in
NGC\,2264. The low disk fraction of these very fast rotators is the
only feature of our sample that could \emph{potentially} be
interpreted as an evidence for disk braking.

However, the lack of evidence for disk braking (in the form of a
correlation between PMS stellar rotation periods and IR excess) in all
but the fastest rotators is not enough to rule out disk braking in PMS
stars.  As shown by \citet{rebull04}, current observational signatures
of disk braking may be hidden by an initial large distribution
of rotation periods in PMS stars.  Because there is currently no
alternative mechanism for angular momentum loss in PMS stars and
because there is evidence for star--disk interaction in very young
stellar objects, a rigorous quantitative analysis of the effects of
disk-braking parameters on current observational signatures is
required to determine whether disk-braking may indeed play a
significant role in the angular momentum evolution of these
stars.  Simulations similar to those run by \citet{rebull04}, with the
improvements suggested in Section~\ref{results} of this paper, can
further constrain the importance of disk braking in the evolution of
PMS stars.  We are in the process of testing such models and will
present our results in a follow-up paper.

\acknowledgments 
We would like to thank the referee for the close reading of the paper.
We thank Luisa Rebull and Deborah Padgett for their detailed suggestions.  
We also thank Paul Harvey, Daniel Jaffe and Neal Evans for their useful 
comments. We thank Judit Ries for obtaining R and I images and standards for
absolute photometry. We also thank Luisa Rebull for providing a
preprint of the paper \citet{rebull05} prior to publication. Support
for this work, part of the Spitzer Legacy Science Program, was
provided by NASA through contract 1224608 issued by the Jet Propulsion
Laboratory, California Institute of Technology, under NASA contract
1407. This publication makes use of data products from the Two Micron
All Sky Survey, which is a joint project of the University of
Massachusetts and the Infrared Processing and Analysis Center funded
by NASA and the National Science Foundation.

%%%%%%%%%%%% TABLE 1 %%%%%%%%%%%%%%%%%%%%%%%%%%%%%
\clearpage
\pagestyle{empty}
\setlength{\voffset}{10mm}
\begin{deluxetable}{rccccccccccccccccccc}
\rotate
\tablewidth{0pt}
\tabletypesize{\tiny}
\tablecaption{Periodic Sample in IC 348\label{xgfp_parameters}}
\tablehead{\colhead{ID}&\colhead{Ra}&\colhead{Dec}&\colhead{P}&\colhead{PS}\tablenotemark{1}&\colhead{$Ref_{P}$\tablenotemark{2}}&\colhead{SpT}&\colhead{$R_{C}$\tablenotemark{3}}&\colhead{$I_{C}$}&\colhead{J}&\colhead{H}&\colhead{$K_{S}$}&\colhead{[3.6]}&\colhead{[3.6] error}&\colhead{[4.5]}&\colhead{[4.5] error}&\colhead{[5.8]}&\colhead{[5.8] error}&\colhead{[8.0]}&\colhead{[8.0] error} \\
\colhead{}  & \multicolumn{2}{c}{(deg)} & \colhead{(days)} &\colhead{} &\colhead{} &\colhead{} & \multicolumn{5}{c}{(mag)}    &\multicolumn{8}{c} {(mJy)}}
\startdata
  1  &   55.5847  &   32.0919  &     7.7  &   44  &        1  &     ---   &    ---   &    ---   &   13.04  &   11.96    &   11.47 	    &  1.18e+01  &  1.00e-01  &  9.33e+00  &  8.22e-02  &  7.48e+00  &  6.35e-02  &  9.03e+00  &  6.65e-02 \\
  2  &   55.6078  &   32.3506  &    17.7  &   61  &        1  &     ---   &   16.74  &   15.30  &   13.29  &   12.32    &   12.00 	    &   ---   &   ---   &   ---   &   ---   &   ---   &   ---   &   ---   &   ---  \\
  3  &   55.6177  &   32.5133  &     4.3  &   47  &        1  &     ---   &   16.78  &   15.25  &   13.18  &   12.31    &   11.99 	    &   ---   &   ---   &   ---   &   ---   &   ---   &   ---   &   ---   &   ---  \\
  4  &   55.6257  &   32.2471  &    26.9  &   24  &        1  &     ---   &   17.29  &   16.30  &   14.64  &   13.88    &   13.62 	    &   ---   &   ---   &   ---   &   ---   &   ---   &   ---   &   ---   &   ---  \\
  5  &   55.6340  &   32.4914  &     2.3  &   30  &        1  &     ---   &   14.51  &   13.60  &   12.33  &   11.67    &   11.45 	    &   ---   &   ---   &   ---   &   ---   &   ---   &   ---   &   ---   &   ---  \\
  6  &   55.6467  &   32.5651  &     0.2  &   35  &        1  &     ---   &   16.12  &   15.42  &   14.63  &   14.24    &   14.03 	    &   ---   &   ---   &   ---   &   ---   &   ---   &   ---   &   ---   &   ---  \\
  7  &   55.6486  &   32.2888  &     1.2  &   43  &        1  &     ---   &   15.89  &   15.02  &   13.84  &   13.14    &   12.95 	    &   ---   &   ---   &   ---   &   ---   &   ---   &   ---   &   ---   &   ---  \\
  8  &   55.6698  &   32.5729  &     5.8  &   45  &        1  &     ---   &    ---   &    ---   &   13.34  &   12.45    &   12.11 	    &   ---   &   ---   &   ---   &   ---   &   ---   &   ---   &   ---   &   ---  \\
  9  &   55.6703  &   32.2263  &     0.5  &   30  &        1  &     ---   &   16.81  &   15.07  &   12.75  &   11.85    &   11.52 	    &  9.55e+00  &  9.10e-02  &   ---   &   ---   &  4.58e+00  &  5.27e-02  &   ---   &   ---  \\
 10  &   55.6817  &   31.9875  &     2.2  &   60  &        1  &     ---   &   15.03  &   13.76  &   11.79  &   10.85    &   10.53 	    &  2.22e+01  &  1.77e-01  &  1.40e+01  &  9.42e-02  &  9.37e+00  &  5.66e-02  &  5.36e+00  &  5.31e-02 \\
 11  &   55.7311  &   31.9534  &     1.9  &   26  &        1  &     ---   &   15.90  &   14.67  &   13.28  &   12.54    &   12.29 	    &  3.95e+00  &  3.21e-02  &  2.56e+00  &  2.26e-02  &  1.79e+00  &  2.79e-02  &  9.82e-01  &  2.73e-02 \\
 12  &   55.7332  &   31.9783  &    22.1  &   29  &        1  &     ---   &    ---   &    ---   &   10.55  &    9.73    &    9.02 	    &  3.84e+02  &  5.69e+00  &  3.23e+02  &  5.02e+00  &  3.69e+02  &  3.62e+00  &  4.88e+02  &  4.60e+00 \\
 13  &   55.7589  &   32.1243  &    17.3  &   31  &        1  &     ---   &   17.20  &   15.30  &   12.83  &   12.05    &   11.69 	    &  8.59e+00  &  1.06e-01  &  6.00e+00  &  6.94e-02  &  3.79e+00  &  5.15e-02  &  2.12e+00  &  3.35e-02 \\
 14  &   55.7794  &   32.1718  &     6.2  &   55  &        1  &     ---   &   16.22  &   14.79  &   12.71  &   11.71    &   11.37 	    &  1.00e+01  &  6.95e-02  &   ---   &   ---   &  4.43e+00  &  4.06e-02  &   ---   &   ---  \\
 15  &   55.7944  &   32.5923  &     6.0  &   34  &        1  &     ---   &   16.44  &   14.98  &   12.98  &   12.04    &   11.71 	    &   ---   &   ---   &   ---   &   ---   &   ---   &   ---   &   ---   &   ---  \\
 16  &   55.7981  &   32.4422  &     7.5  &   23  &        1  &     ---   &   16.79  &   15.79  &   14.33  &   13.57    &   13.35 	    &  1.52e+00  &  2.52e-02  &  1.01e+00  &  1.39e-02  &  6.56e-01  &  2.93e-02  &  3.35e-01  &  3.20e-02 \\
 17  &   55.8014  &   32.5698  &     0.2  &   23  &        1  &     ---   &   16.14  &   15.20  &   14.35  &   13.75    &   13.56 	    &   ---   &   ---   &   ---   &   ---   &   ---   &   ---   &   ---   &   ---  \\
 18  &   55.8072  &   32.0125  &     7.8  &   24  &        1  &     ---   &   23.28  &   19.96  &   14.26  &   12.57    &   11.76 	    &  1.26e+01  &  1.02e-01  &  1.12e+01  &  9.15e-02  &  1.10e+01  &  8.25e-02  &  1.32e+01  &  9.20e-02 \\
 19  &   55.8482  &   32.2072  &     3.0  &   36  &        1  &     ---   &   18.84  &   16.79  &   14.09  &   13.15    &   12.72 	    &  4.16e+00  &  5.29e-02  &  3.26e+00  &  3.85e-02  &  2.82e+00  &  3.71e-02  &  2.30e+00  &  3.73e-02 \\
 20  &   55.8516  &   32.6421  &     1.3  &   54  &        1  &     ---   &   16.13  &   14.56  &   12.25  &   11.28    &   10.92 	    &   ---   &   ---   &   ---   &   ---   &   ---   &   ---   &   ---   &   ---  \\
 21  &   55.8675  &   32.0331  &     8.8  &   30  &        1  &     ---   &   14.94  &   13.61  &   11.76  &   10.72    &   10.11 	    &  4.45e+01  &  9.46e-01  &  5.06e+01  &  5.65e-01  &  3.85e+01  &  2.95e-01  &  3.35e+01  &  2.40e-01 \\
 22  &   55.8875  &   32.4674  &    12.2  &   27  &        1  &     ---   &   17.47  &   15.81  &   13.44  &   12.42    &   12.08 	    &  5.32e+00  &  1.21e-01  &  3.52e+00  &  3.59e-02  &  2.54e+00  &  7.28e-02  &  1.31e+00  &  4.10e-02 \\
 23  &   55.9382  &   32.0663  &    27.6  &   52  &        1  &     ---   &   19.29  &   17.70  &   13.78  &   12.15    &   11.01 	    &  2.35e+01  &  5.18e+00  &  1.45e+02  &  2.57e+00  &  6.61e+01  &  1.54e+00  &  1.03e+02  &  2.04e+00 \\
 24  &   55.9495  &   32.2991  &     9.8  &   54  &        1  &     ---   &   15.79  &   14.44  &   12.59  &   11.68    &   11.39 	    &  9.19e+00  &  1.46e-01  &  6.21e+00  &  7.46e-02  &  4.28e+00  &  4.46e-02  &  2.48e+00  &  3.31e-02 \\
 25  &   55.9532  &   32.1259  &    20.3  &   27  &        1  &     M1.5  &   15.44  &   14.16  &   12.43  &   11.65    &   11.37 	    &  9.01e+00  &  1.79e-01  &  6.60e+00  &  7.88e-02  &  4.12e+00  &  5.04e-02  &  2.31e+00  &  4.18e-02 \\
 26  &   55.9534  &   32.2643  &     2.9  &   46  &        1  &     ---   &   17.02  &   15.32  &   13.01  &   12.14    &   11.78 	    &  8.50e+00  &  1.52e-01  &  8.54e+00  &  9.87e-02  &  8.40e+00  &  7.56e-02  &  1.04e+01  &  7.31e-02 \\
 27  &   55.9558  &   32.1777  &    13.1  &   44  &        1  &     M3.5  &   16.36  &   14.76  &   12.68  &   11.85    &   11.56 	    &  7.51e+00  &  1.14e-01  &  5.42e+00  &  6.95e-02  &  3.55e+00  &  4.44e-02  &  2.12e+00  &  4.02e-02 \\
 28  &   55.9642  &   32.5302  &     3.8  &   60  &        1  &     ---   &   15.62  &   14.18  &   12.03  &   11.12    &   10.80 	    &   ---   &   ---   &   ---   &   ---   &   ---   &   ---   &   ---   &   ---  \\
 29  &   55.9802  &   31.9256  &     9.6  &   26  &        1  &     ---   &   20.92  &   18.30  &   14.22  &   12.31    &   11.39 	    &  2.35e+01  &  4.41e-01  &  2.18e+01  &  2.00e-01  &  1.80e+01  &  1.42e-01  &  1.19e+01  &  8.50e-02 \\
 30  &   55.9843  &   32.5050  &     8.1  &   52  &        1  &     ---   &   15.89  &   14.51  &   12.49  &   11.62    &   11.33 	    &   ---   &   ---   &   ---   &   ---   &   ---   &   ---   &   ---   &   ---  \\
 31  &   55.9940  &   32.2910  &    12.8  &   26  &        1  &     ---   &   15.85  &   14.36  &   12.05  &   11.12    &   10.66 	    &  3.06e+01  &  4.63e-01  &  2.18e+01  &  4.05e-01  &  2.06e+01  &  1.60e-01  &  2.43e+01  &  1.70e-01 \\
 32  &   55.9962  &   32.2392  &    17.7  &   28  &        1  &     ---   &   17.27  &   15.82  &   13.49  &   12.29    &   11.39 	    &  1.31e+01  &  2.03e-01  &  1.40e+01  &  1.76e-01  &  1.19e+01  &  9.46e-02  &  1.26e+01  &  9.91e-02 \\
 33  &   55.9981  &   32.2653  &     7.6  &   39  &        1  &     ---   &   16.96  &   15.43  &   13.06  &   11.99    &   11.69 	    &  7.19e+00  &  1.24e-01  &  4.91e+00  &  5.29e-02  &  3.42e+00  &  4.47e-02  &  1.94e+00  &  3.30e-02 \\
 34  &   55.9988  &   32.2341  &    14.0  &   63  &      1,2  &    M0.75  &   15.50  &   14.19  &   12.30  &   11.40    &   11.06 	    &  1.29e+01  &  1.97e-01  &  8.97e+00  &  9.60e-02  &  6.17e+00  &  6.52e-02  &  3.43e+00  &  5.81e-02 \\
 35  &   56.0090  &   32.3278  &     6.2  &   63  &        1  &     ---   &   15.73  &   14.41  &   12.51  &   11.64    &   11.38 	    &  8.84e+00  &  1.47e-01  &  6.48e+00  &  7.87e-02  &  4.48e+00  &  3.95e-02  &  2.62e+00  &  3.43e-02 \\
 36  &   56.0177  &   32.2305  &     1.2  &   32  &        1  &    M4.75  &    ---   &    ---   &   12.61  &   11.76    &   11.41 	    &  9.73e+00  &  1.41e-01  &  7.41e+00  &  8.13e-02  &  4.96e+00  &  4.95e-02  &  3.05e+00  &  5.64e-02 \\
 37  &   56.0283  &   32.1317  &     1.3  &   23  &        1  &    M4.25  &   17.27  &   15.48  &   13.03  &   12.14    &   11.75 	    &  1.01e+01  &  1.73e-01  &  8.91e+00  &  1.05e-01  &  7.53e+00  &  8.79e-02  &  8.11e+00  &  1.05e-01 \\
 38  &   56.0422  &   32.0679  &     2.2  &   --  &        3  &    M5.75  &   18.25  &   16.14  &   13.22  &   12.41    &   11.93 	    &  8.72e+00  &  1.14e-01  &  8.23e+00  &  8.07e-02  &  7.25e+00  &  1.17e-01  &  7.28e+00  &  1.22e-01 \\
 39  &   56.0468  &   32.1378  &    13.0  &   --  &        3  &    M5.25  &   17.38  &   15.39  &   12.84  &   12.09    &   11.75 	    &  7.36e+00  &  1.05e-01  &  5.35e+00  &  6.98e-02  &  3.72e+00  &  5.52e-02  &  2.27e+00  &  6.96e-02 \\
 40  &   56.0469  &   32.1034  &     9.1  &   29  &  1,2,3,4  &       M0  &   15.72  &   14.34  &   12.42  &   11.44    &   11.16 	    &  1.12e+01  &  1.41e-01  &  7.35e+00  &  8.22e-02  &  5.19e+00  &  7.27e-02  &  2.75e+00  &  7.99e-02 \\
 41  &   56.0476  &   32.3278  &     0.8  &   42  &      1,2  &       M3  &   16.13  &   14.65  &   12.56  &   11.78    &   11.51 	    &  8.94e+00  &  1.32e-01  &  6.19e+00  &  6.99e-02  &  4.51e+00  &  4.54e-02  &  2.42e+00  &  3.28e-02 \\
 42  &   56.0574  &   31.9263  &     8.3  &   39  &        1  &     ---   &   19.31  &   17.13  &   13.68  &   12.19    &   11.64 	    &  1.14e+01  &  2.12e-01  &  7.98e+00  &  2.19e-01  &  5.13e+00  &  6.85e-02  &  3.64e+00  &  9.47e-02 \\
 43  &   56.0649  &   32.1561  &     0.6  &   --  &       3,  &     M7.5  &   20.94  &   18.18  &   14.59  &   13.78    &   13.30 	    &  2.11e+00  &  2.42e-02  &  1.52e+00  &  1.78e-02  &  1.11e-01  &  5.27e-02  &  4.77e-01  &  1.96e-02 \\
 44  &   56.0654  &   32.5245  &    13.7  &   44  &        1  &     ---   &   14.21  &   13.07  &   11.72  &   10.98    &   10.75 	    &   ---   &   ---   &   ---   &   ---   &   ---   &   ---   &   ---   &   ---  \\
 45  &   56.0684  &   32.1653  &     3.0  &   33  &  1,2,3,4  &       K0  &    ---   &    ---   &   11.32  &   10.58    &   10.38 	    &  2.06e+01  &  2.90e-01  &  1.36e+01  &  1.54e-01  &  9.46e+00  &  8.43e-02  &  5.40e+00  &  7.21e-02 \\
 46  &   56.0746  &   32.2056  &     4.5  &   29  &        1  &     M2.5  &   15.59  &   14.10  &   12.16  &   11.35    &   11.07 	    &  1.17e+01  &  1.86e-01  &  8.08e+00  &  9.06e-02  &  5.55e+00  &  5.84e-02  &  3.06e+00  &  5.62e-02 \\
 47  &   56.0758  &   32.1665  &     2.7  &   27  &        1  &    M4.25  &   17.87  &   15.85  &   13.21  &   12.27    &   11.87 	    &  7.07e+00  &  1.09e-01  &  6.09e+00  &  7.55e-02  &  4.88e+00  &  5.76e-02  &  4.04e+00  &  6.86e-02 \\
 48  &   56.0761  &   32.1257  &     2.2  &   --  &        3  &    M4.75  &   18.27  &   16.30  &    ---   &    ---     &    ---  	    &  3.74e+00  &  4.99e-02  &  2.79e+00  &  3.65e-02  &  1.94e+00  &  4.44e-02  &  2.00e+00  &  5.69e-02 \\
 49  &   56.0801  &   32.1262  &     7.6  &   --  &        4  &    M3.75  &    ---   &    ---   &    ---   &    ---     &    ---  	    &  1.50e+01  &  2.29e-01  &  1.28e+01  &  1.40e-01  &  1.16e+01  &  1.53e-01  &  1.28e+01  &  1.42e-01 \\
 50  &   56.0834  &   32.1127  &     8.6  &   --  &      3,4  &     M3.5  &   17.77  &   16.04  &   13.58  &   12.61    &   12.23 	    &  4.66e+00  &  6.74e-02  &  3.41e+00  &  3.37e-02  &  2.18e+00  &  4.54e-02  &  8.11e-01  &  3.81e-02 \\
 51  &   56.0841  &   32.1491  &     2.2  &   --  &        3  &       M2  &   16.76  &   15.03  &   12.70  &   11.79    &   11.41 	    &  1.01e+01  &  1.69e-01  &  9.42e+00  &  1.03e-01  &  7.79e+00  &  9.73e-02  &  7.59e+00  &  1.51e-01 \\
 52  &   56.0843  &   32.5063  &     2.8  &   45  &        1  &     ---   &   17.11  &   15.42  &   13.13  &   12.20    &   11.85 	    &  7.23e+00  &  7.55e-02  &  4.77e+00  &  7.92e-02  &  3.20e+00  &  4.29e-02  &  1.81e+00  &  5.97e-02 \\
 53  &   56.0857  &   32.4610  &     4.8  &   51  &        1  &     ---   &   17.66  &   15.79  &   13.23  &   12.37    &   12.01 	    &  6.86e+00  &  9.46e-02  &  4.49e+00  &  5.25e-02  &  2.87e+00  &  5.34e-02  &  1.54e+00  &  4.43e-02 \\
 54  &   56.0886  &   32.0840  &     6.9  &   54  &      1,3  &     M2.5  &   16.82  &   15.17  &   12.74  &   11.73    &   11.40 	    &  9.64e+00  &  1.35e-01  &  7.12e+00  &  7.44e-02  &  4.79e+00  &  8.04e-02  &  2.20e+00  &  7.35e-02 \\
 55  &   56.0886  &   32.2103  &     2.3  &   30  &        1  &    M4.75  &   17.41  &   15.71  &   13.70  &   12.92    &   12.51 	    &  4.90e+00  &  5.86e-02  &  4.30e+00  &  3.88e-02  &  3.50e+00  &  3.99e-02  &  2.71e+00  &  6.23e-02 \\
 56  &   56.0898  &   32.1715  &     7.0  &   60  &  1,2,3,4  &     M1.5  &   16.20  &   14.71  &   12.62  &   11.66    &   11.35 	    &  1.02e+01  &  1.13e-01  &  7.13e+00  &  6.83e-02  &  4.88e+00  &  5.12e-02  &  2.68e+00  &  3.12e-02 \\
 57  &   56.0903  &   32.1069  &     8.4  &   --  &     2,4,  &    M2.75  &   16.21  &   14.65  &   12.54  &   11.60    &   11.31 	    &  9.91e+00  &  1.40e-01  &  6.67e+00  &  8.79e-02  &  4.74e+00  &  6.77e-02  &  2.72e+00  &  6.81e-02 \\
 58  &   56.0901  &   32.1771  &     2.8  &   26  &      1,2  &       K7  &   16.42  &   14.92  &   12.49  &   11.28    &   10.62 	    &  5.32e+01  &  5.84e-01  &  4.80e+01  &  5.73e-01  &  4.16e+01  &  2.42e-01  &  3.53e+01  &  2.36e-01 \\
 59  &   56.0913  &   32.2032  &    14.0  &   --  &        3  &       M4  &   16.37  &   14.57  &   12.28  &   11.40    &   11.09 	    &  1.43e+01  &  2.09e-01  &  8.91e+00  &  1.13e-01  &  6.13e+00  &  6.78e-02  &  4.04e+00  &  1.10e-01 \\
 60  &   56.0929  &   32.0952  &    30.0  &   25  &        1  &       K8  &   15.88  &   14.45  &   12.55  &   11.28    &   10.70 	    &  2.76e+01  &  3.23e-01  &  2.34e+01  &  2.51e-01  &  2.06e+01  &  1.54e-01  &  2.83e+01  &  2.95e-01 \\
 61  &   56.0930  &   32.2002  &     8.4  &   --  &        3  &       M1  &   16.35  &   14.78  &   12.56  &   11.60    &   11.10 	    &  1.52e+01  &  2.96e-01  &  1.66e+01  &  1.93e-01  &  1.67e+01  &  1.41e-01  &  1.90e+01  &  2.11e-01 \\
 62  &   56.0941  &   32.0316  &     1.0  &   25  &        1  &     M2.5  &   16.01  &   14.40  &   12.12  &   11.15    &   10.79 	    &  1.35e+01  &  2.91e-01  &  1.10e+01  &  1.44e-01  &  8.44e+00  &  6.38e-02  &  6.76e+00  &  6.39e-02 \\
 63  &   56.0982  &   32.1594  &     1.7  &   --  &        3  &       M5  &   17.72  &   15.89  &   13.50  &   12.74    &   12.40 	    &  4.81e+00  &  5.23e-02  &  4.13e+00  &  3.72e-02  &  3.10e+00  &  4.29e-02  &  2.92e+00  &  7.73e-02 \\
 64  &   56.0986  &   32.1129  &    10.0  &   48  &  1,2,3,4  &     M2.5  &   15.83  &   14.27  &   12.23  &   11.38    &   11.07 	    &  1.30e+01  &  1.44e-01  &  8.89e+00  &  6.93e-02  &  6.19e+00  &  6.60e-02  &  3.27e+00  &  5.27e-02 \\
 65  &   56.1024  &   32.0659  &     4.9  &   57  &    1,3,4  &       M1  &   17.29  &   15.54  &   12.82  &   11.81    &   11.37 	    &  9.91e+00  &  1.41e-01  &  6.82e+00  &  7.41e-02  &  4.64e+00  &  7.81e-02  &   ---   &   ---  \\
 66  &   56.1065  &   32.1048  &     7.3  &   52  &      1,3  &    M2.25  &   17.85  &   15.98  &   12.98  &   11.70    &   11.14 	    &  1.55e+01  &  1.75e-01  &  1.25e+01  &  1.22e-01  &  9.82e+00  &  1.03e-01  &  1.05e+01  &  2.08e-01 \\
 67  &   56.1065  &   32.1919  &     5.4  &   24  &    1,2,3  &       M0  &   16.47  &   14.85  &   12.48  &   11.27    &   10.64 	    &  4.36e+01  &  7.81e-01  &  4.11e+01  &  7.02e-01  &  3.58e+01  &  3.29e-01  &  4.81e+01  &  3.75e-01 \\
 68  &   56.1066  &   32.2083  &     8.4  &   63  &    1,2,4  &     M0.5  &   14.79  &   13.59  &   11.82  &   10.95    &   10.68 	    &  1.68e+01  &  2.56e-01  &  1.11e+01  &  1.34e-01  &  7.55e+00  &  7.55e-02  &  4.49e+00  &  6.18e-02 \\
 69  &   56.1110  &   32.0662  &     3.1  &   30  &  1,2,3,4  &    M4.75  &   16.20  &   14.33  &   11.80  &   10.94    &   10.59 	    &  1.93e+01  &  2.23e-01  &  1.26e+01  &  1.49e-01  &  8.79e+00  &  8.03e-02  &  5.22e+00  &  1.33e-01 \\
 70  &   56.1112  &   32.1390  &     8.9  &   29  &        1  &     M0.5  &   17.72  &   16.03  &   12.99  &   11.56    &   10.84 	    &  3.12e+01  &  4.26e-01  &  2.76e+01  &  3.01e-01  &  1.99e+01  &  1.94e-01  &  1.95e+01  &  2.54e-01 \\
 71  &   56.1126  &   32.0788  &     9.1  &   --  &        3  &       M1  &   15.07  &   13.77  &   11.95  &   11.14    &   10.85 	    &  1.44e+01  &  1.80e-01  &  9.96e+00  &  1.49e-01  &  6.82e+00  &  7.61e-02  &  3.88e+00  &  1.06e-01 \\
 72  &   56.1137  &   32.1216  &     1.5  &   --  &        3  &    M4.75  &   18.57  &   16.81  &   14.43  &   13.59    &   13.27 	    &  2.03e+00  &  2.77e-02  &  1.39e+00  &  1.76e-02  &  9.29e-01  &  2.54e-02  &  7.75e-01  &  5.00e-02 \\
 73  &   56.1153  &   32.5638  &     2.6  &   31  &        1  &     ---   &   15.87  &   14.34  &   12.10  &   11.23    &   10.87 	    &   ---   &   ---   &   ---   &   ---   &   ---   &   ---   &   ---   &   ---  \\
 74  &   56.1162  &   32.1255  &     5.4  &   23  &    1,2,4  &       M2  &   15.80  &   14.23  &   12.13  &   11.27    &   10.97 	    &  1.40e+01  &  2.01e-01  &  9.78e+00  &  9.15e-02  &  6.30e+00  &  9.11e-02  &  3.39e+00  &  8.18e-02 \\
 75  &   56.1172  &   32.2667  &     2.7  &   52  &        1  &    M3.25  &   15.98  &   14.38  &   12.21  &   11.35    &   11.02 	    &  1.29e+01  &  1.97e-01  &  8.79e+00  &  1.10e-01  &  6.35e+00  &  7.15e-02  &  3.71e+00  &  8.04e-02 \\
 76  &   56.1186  &   32.1229  &     7.0  &   52  &  1,2,3,4  &     K6.5  &   14.44  &   13.30  &   11.67  &   10.85    &   10.58 	    &  1.65e+01  &  2.50e-01  &  1.12e+01  &  1.14e-01  &  7.29e+00  &  1.16e-01  &  4.04e+00  &  1.01e-01 \\
 77  &   56.1309  &   32.1915  &     1.4  &   --  &        3  &    M5.25  &   18.27  &   16.15  &   13.48  &   12.74    &   12.34 	    &  4.68e+00  &  7.17e-02  &  3.60e+00  &  7.48e-02  &  2.22e+00  &  9.20e-02  &  1.49e+00  &  1.70e-01 \\
 78  &   56.1314  &   32.1458  &    10.8  &   --  &      3,4  &       K2  &    ---   &    ---   &   10.69  &    9.97    &    9.72 	    &  3.89e+01  &  6.15e-01  &  2.66e+01  &  2.96e-01  &  1.80e+01  &  1.46e-01  &  1.05e+01  &  2.07e-01 \\
 79  &   56.1349  &   32.0576  &     1.6  &   --  &        3  &     M5.5  &   20.68  &   18.09  &   14.88  &   14.04    &   13.48 	    &  2.37e+00  &  2.96e-02  &  2.05e+00  &  2.29e-02  &  1.63e+00  &  6.07e-02  &   ---   &   ---  \\
 80  &   56.1357  &   32.1488  &     6.7  &   --  &        3  &       M3  &   16.05  &   14.43  &   12.11  &   11.13    &   10.76 	    &  1.77e+01  &  2.94e-01  &  1.40e+01  &  1.39e-01  &  1.09e+01  &  1.30e-01  &  9.29e+00  &  1.65e-01 \\
 81  &   56.1364  &   32.1437  &     2.6  &   --  &      3,4  &       G6  &    ---   &    ---   &   10.28  &    9.65    &    9.43 	    &  4.86e+01  &  6.93e-01  &  3.47e+01  &  2.77e-01  &  2.25e+01  &  2.15e-01  &  1.26e+01  &  2.76e-01 \\
 82  &   56.1365  &   32.1544  &     5.5  &   33  &      1,4  &    M3.25  &   16.14  &   14.68  &   12.35  &   11.37    &   11.01 	    &  1.39e+01  &  1.80e-01  &  1.00e+01  &  1.22e-01  &  7.12e+00  &  8.31e-02  &  4.88e+00  &  1.40e-01 \\
 83  &   56.1367  &   32.0704  &     5.3  &   23  &        1  &       M5  &   18.61  &   16.49  &   13.74  &   12.90    &   12.52 	    &  4.15e+00  &  5.86e-02  &  2.86e+00  &  4.19e-02  &  1.90e+00  &  5.41e-02  &   ---   &   ---  \\
 84  &   56.1388  &   32.1610  &     2.2  &   25  &        1  &       M2  &   15.10  &   13.99  &   12.33  &   11.37    &   11.06 	    &  2.13e+01  &  3.10e-01  &  1.62e+01  &  2.98e-01  &  1.20e+01  &  1.46e-01  &  7.49e+00  &  1.75e-01 \\
 85  &   56.1408  &   31.9751  &     3.9  &   27  &        1  &     ---   &   18.13  &   16.23  &   13.49  &   12.27    &   11.61 	    &  1.13e+01  &  1.95e-01  &  9.15e+00  &  9.60e-02  &  6.87e+00  &  8.20e-02  &  6.63e+00  &  5.42e-02 \\
 86  &   56.1416  &   32.1484  &    16.4  &   --  &      2,4  &       M0  &    ---   &    ---   &   11.85  &   10.98    &   10.70 	    &  1.58e+01  &  2.32e-01  &  1.06e+01  &  1.18e-01  &  7.62e+00  &  1.00e-01  &  5.25e+00  &  1.65e-01 \\
 87  &   56.1419  &   32.1159  &     3.4  &   --  &        3  &    M7.25  &   19.47  &   17.26  &    ---   &    ---     &    ---  	    &  4.67e+00  &  6.25e-02  &  4.28e+00  &  4.42e-02  &  4.28e+00  &  6.56e-02  &  4.86e+00  &  1.10e-01 \\
 88  &   56.1450  &   31.9487  &     3.5  &   35  &        1  &     ---   &   16.82  &   15.17  &   13.03  &   12.09    &   11.62 	    &  1.26e+01  &  1.21e-01  &  8.60e+00  &  1.18e-01  &  6.43e+00  &  6.74e-02  &  5.19e+00  &  1.12e-01 \\
 89  &   56.1453  &   32.1094  &     5.4  &   --  &    2,3,4  &     K5.5  &   14.55  &   13.35  &   11.51  &   10.61    &   10.31 	    &  2.32e+01  &  3.68e-01  &  1.63e+01  &  1.91e-01  &  1.13e+01  &  1.30e-01  &  6.64e+00  &  1.27e-01 \\
 90  &   56.1459  &   32.1493  &     1.8  &   --  &        3  &    M4.75  &   18.08  &   16.63  &   14.87  &   14.11    &   13.71 	    &  1.79e+00  &  4.59e-02  &  1.32e+00  &  3.15e-02  &  1.01e+00  &  1.04e-01  &  5.88e-01  &  2.12e-01 \\
 91  &   56.1460  &   32.1270  &     4.5  &   42  &      1,2  &     K6.5  &   14.22  &   15.14  &   10.99  &   10.07    &    9.76 	    &  4.18e+01  &  6.16e-01  &  2.80e+01  &  3.81e-01  &  1.81e+01  &  1.49e-01  &  1.17e+01  &  1.38e-01 \\
 92  &   56.1477  &   32.1490  &     1.9  &   --  &        3  &    M5.25  &   17.71  &   16.05  &   13.25  &   12.31    &   11.83 	    &  9.19e+00  &  1.78e-01  &  8.21e+00  &  1.22e-01  &  7.54e+00  &  1.48e-01  &  7.48e+00  &  1.86e-01 \\
 93  &   56.1480  &   32.1346  &     1.7  &   --  &        3  &    M5.25  &   18.26  &   16.39  &   13.64  &   12.72    &   12.33 	    &  4.72e+00  &  7.17e-02  &  3.38e+00  &  4.57e-02  &  2.35e+00  &  3.62e-02  &  1.14e+00  &  9.20e-02 \\
 94  &   56.1487  &   32.0510  &    12.0  &   34  &        1  &    M3.25  &   19.06  &   17.36  &   14.13  &   12.78    &   11.84 	    &  1.15e+01  &  1.96e-01  &  1.18e+01  &  1.63e-01  &  1.16e+01  &  9.46e-02  &  1.30e+01  &  1.35e-01 \\
 95  &   56.1539  &   32.1126  &     1.7  &   --  &      2,4  &       G3  &    ---   &    ---   &    9.21  &    8.48    &    8.19 	    &  1.85e+02  &  3.63e+00  &  1.03e+02  &  2.98e+00  &  1.15e+02  &  7.00e-01  &  8.48e+01  &  6.24e-01 \\
 96  &   56.1541  &   32.1429  &     2.5  &   --  &        3  &    M4.75  &   17.53  &   15.73  &   13.06  &   12.07    &   11.59 	    &  1.05e+01  &  1.62e-01  &  8.46e+00  &  1.08e-01  &  6.92e+00  &  1.27e-01  &  6.27e+00  &  1.59e-01 \\
 97  &   56.1558  &   32.1033  &     6.2  &   --  &      2,4  &       K7  &   15.17  &   13.88  &   12.08  &   11.14    &   10.87 	    &  1.41e+01  &  2.05e-01  &  9.42e+00  &  1.10e-01  &  6.86e+00  &  8.12e-02  &  3.96e+00  &  1.28e-01 \\
 98  &   56.1558  &   32.2067  &    19.8  &   31  &      1,3  &       M2  &   17.89  &   16.02  &   13.07  &   11.84    &   11.28 	    &  1.32e+01  &  2.18e-01  &  1.15e+01  &  1.27e-01  &  8.93e+00  &  1.04e-01  &  1.15e+01  &  2.06e-01 \\
 99  &   56.1559  &   32.1502  &     8.4  &   38  &      1,4  &       M1  &   16.28  &   14.91  &   12.48  &   11.44    &   10.99 	    &  2.25e+01  &  4.00e-01  &  2.04e+01  &  3.67e-01  &  1.89e+01  &  2.91e-01  &  2.39e+01  &  3.96e-01 \\
100  &   56.1574  &   32.2050  &     3.3  &   58  &    1,3,4  &     M4.5  &   17.69  &   15.68  &   13.04  &   12.15    &   11.79 	    &  8.40e+00  &  8.81e-02  &  5.95e+00  &  5.99e-02  &  4.12e+00  &  5.34e-02  &  1.97e+00  &  4.63e-02 \\
101  &   56.1578  &   32.1345  &     8.2  &   30  &      1,3  &       K7  &   15.58  &   14.12  &   11.69  &   10.48    &    9.83 	    &  6.03e+01  &  1.51e+00  &  6.55e+01  &  8.83e-01  &  5.62e+01  &  4.69e-01  &  5.83e+01  &  5.54e-01 \\
102  &   56.1583  &   32.0582  &     8.6  &   33  &      1,2  &       K6  &   14.66  &   13.35  &   11.45  &   10.44    &    9.87 	    &  6.15e+01  &  6.67e-01  &  4.52e+01  &  8.42e-01  &  4.40e+01  &  3.47e-01  &  5.27e+01  &  4.02e-01 \\
103  &   56.1599  &   32.2166  &    13.5  &   --  &        4  &       M0  &   15.97  &   14.64  &   12.81  &   11.93    &   11.65 	    &  7.73e+00  &  1.08e-01  &  4.99e+00  &  4.89e-02  &  3.22e+00  &  4.41e-02  &  1.52e+00  &  6.94e-02 \\
104  &   56.1602  &   32.1266  &     5.1  &   45  &  1,2,3,4  &       K6  &   14.52  &   13.21  &   11.18  &   10.23    &    9.85 	    &  3.77e+01  &  8.08e-01  &  3.67e+01  &  4.80e-01  &  3.81e+01  &  2.85e-01  &  5.69e+01  &  5.22e-01 \\
105  &   56.1606  &   32.1335  &     7.3  &   29  &  1,2,3,4  &    M1.25  &   15.85  &   14.32  &   11.94  &   10.90    &   10.47 	    &  2.74e+01  &  4.85e-01  &  2.45e+01  &  2.55e-01  &  1.88e+01  &  2.27e-01  &  2.65e+01  &  3.43e-01 \\
106  &   56.1612  &   32.1491  &     2.1  &   --  &        3  &    M3.25  &   15.99  &   14.45  &   12.45  &   11.57    &   11.28 	    &  1.18e+01  &  1.61e-01  &  7.93e+00  &  8.67e-02  &  5.09e+00  &  1.39e-01  &  3.13e+00  &  1.50e-01 \\
107  &   56.1613  &   32.1450  &     2.4  &   31  &    1,3,4  &       K3  &   15.60  &   14.01  &   11.19  &    9.97    &    9.49 	    &  5.52e+01  &  9.19e-01  &  4.12e+01  &  3.49e-01  &  2.72e+01  &  2.71e-01  &  1.50e+01  &  2.52e-01 \\
108  &   56.1616  &   32.3182  &     4.8  &   56  &        1  &     ---   &   17.16  &   15.47  &   13.34  &   12.47    &   12.14 	    &  5.25e+00  &  8.46e-02  &  3.49e+00  &  5.26e-02  &  2.62e+00  &  7.57e-02  &  1.79e+00  &  1.08e-01 \\
109  &   56.1632  &   32.1551  &     1.6  &   --  &    2,3,4  &       G8  &    ---   &    ---   &   10.07  &    9.14    &    8.77 	    &  9.01e+01  &  2.24e+00  &  7.49e+01  &  9.51e-01  &  5.02e+01  &  5.20e-01  &  2.87e+01  &  6.57e-01 \\
110  &   56.1633  &   32.1625  &     3.9  &   52  &    1,3,4  &       M2  &   16.30  &   14.77  &   12.46  &   11.46    &   11.09 	    &  1.66e+01  &  2.69e-01  &  1.48e+01  &  1.71e-01  &  1.27e+01  &  1.47e-01  &  1.41e+01  &  3.25e-01 \\
111  &   56.1643  &   32.1689  &     1.5  &   --  &        3  &       M5  &   17.60  &   16.06  &   13.85  &   13.10    &   12.72 	    &  3.16e+00  &  4.50e-02  &  2.33e+00  &  3.03e-02  &  1.37e+00  &  4.53e-02  &  1.28e+00  &  1.37e-01 \\
112  &   56.1638  &   32.6037  &     0.2  &   24  &        1  &     ---   &   19.75  &   18.64  &   16.28  &   15.33    &   15.12 	    &   ---   &   ---   &   ---   &   ---   &   ---   &   ---   &   ---   &   ---  \\
113  &   56.1658  &   32.3012  &     9.5  &   38  &        1  &    M3.75  &   16.85  &   15.00  &   12.23  &   11.28    &   10.78 	    &  2.21e+01  &  3.09e-01  &  1.75e+01  &  2.43e-01  &  1.37e+01  &  1.10e-01  &  1.35e+01  &  1.05e-01 \\
114  &   56.1692  &   32.3864  &     4.1  &   45  &        1  &     ---   &    ---   &    ---   &   12.80  &   11.83    &   11.55 	    &  8.07e+00  &  1.24e-01  &  5.61e+00  &  5.73e-02  &  3.69e+00  &  4.76e-02  &  2.16e+00  &  3.49e-02 \\
115  &   56.1721  &   32.1737  &     3.7  &   --  &        3  &    M4.75  &   16.05  &   14.76  &   12.56  &   11.77    &   11.44 	    &  9.64e+00  &  1.52e-01  &  7.34e+00  &  7.02e-02  &  4.75e+00  &  7.29e-02  &  1.86e+00  &  7.88e-02 \\
116  &   56.1721  &   32.0815  &     1.6  &   --  &        3  &       M5  &    ---   &    ---   &   14.51  &   13.71    &   13.28 	    &  1.95e+00  &  2.96e-02  &  1.43e+00  &  1.77e-02  &  8.46e-01  &  3.85e-02  &   ---   &   ---  \\
117  &   56.1732  &   32.1776  &     1.7  &   --  &        3  &    M5.75  &   19.07  &   18.16  &   15.46  &   14.86    &   14.25 	    &  1.02e+00  &  1.70e-02  &  7.93e-01  &  1.10e-02  &  5.05e-01  &  4.12e-02  &  3.60e-01  &  2.61e-02 \\
118  &   56.1739  &   32.2006  &     2.1  &   28  &        1  &       M5  &   18.68  &   16.09  &   13.26  &   12.02    &   11.42 	    &  1.12e+01  &  1.16e-01  &  8.13e+00  &  7.94e-02  &  6.70e+00  &  7.23e-02  &  6.61e+00  &  1.46e-01 \\
119  &   56.1753  &   32.1503  &    16.0  &   38  &        1  &     ---   &   16.27  &   14.21  &   11.79  &   10.67    &   10.13 	    &  2.47e+01  &  3.83e-01  &  2.21e+01  &  4.64e-01  &  2.61e+01  &  4.01e-01  &  2.10e+01  &  3.84e-01 \\
120  &   56.1774  &   32.1674  &     3.6  &   --  &        3  &    M4.25  &   19.18  &   17.11  &   13.63  &   12.36    &   11.79 	    &  1.18e+01  &  1.33e-01  &  1.12e+01  &  1.02e-01  &  9.82e+00  &  9.29e-02  &  1.02e+01  &  1.64e-01 \\
121  &   56.1776  &   32.1054  &    12.0  &   25  &  1,2,3,4  &       M1  &   15.10  &   13.95  &   12.52  &   11.77    &   11.54 	    &  8.41e+00  &  1.09e-01  &  5.26e+00  &  5.70e-02  &  3.51e+00  &  8.74e-02  &  2.64e+00  &  1.48e-01 \\
122  &   56.1791  &   32.0259  &     1.0  &   24  &        1  &     ---   &    ---   &    ---   &   15.22  &   14.66    &   11.74 	    &   ---   &   ---   &   ---   &   ---   &   ---   &  4.39e-01  &   ---   &   ---  \\
123  &   56.1824  &   32.1751  &    10.6  &   26  &      1,2  &    M1.25  &   15.92  &   14.40  &   12.29  &   11.25    &   10.83 	    &  1.82e+01  &  2.31e-01  &  1.85e+01  &  1.77e-01  &  2.08e+01  &  1.29e-01  &  3.27e+01  &  2.58e-01 \\
124  &   56.1843  &   32.1465  &     1.8  &   --  &        3  &    M5.75  &   19.20  &   17.42  &   14.59  &   13.71    &   13.34 	    &  2.21e+00  &  2.77e-02  &  1.49e+00  &  1.78e-02  &  5.70e-01  &  4.44e-02  &   ---   &  8.85e-02 \\
125  &   56.1845  &   32.1769  &     1.5  &   --  &        3  &    M5.25  &   20.34  &   18.00  &   15.38  &   14.75    &   14.37 	    &  8.78e-01  &  1.18e-02  &  6.42e-01  &  1.35e-02  &  3.74e-01  &  3.87e-02  &   ---   &  6.35e-02 \\
126  &   56.1902  &   32.1863  &     1.3  &   --  &        3  &    M4.75  &   19.33  &   17.46  &   15.10  &   14.29    &   13.95 	    &  1.14e+00  &  2.28e-02  &  8.23e-01  &  1.39e-02  &  5.11e-01  &  4.84e-02  &  3.04e-01  &  1.36e-01 \\
127  &   56.2034  &   32.2227  &     6.9  &   61  &        1  &    M2.75  &   16.80  &   15.20  &   13.10  &   12.24    &   11.94 	    &  5.62e+00  &  7.73e-02  &  3.86e+00  &  3.98e-02  &  2.63e+00  &  5.02e-02  &  1.57e+00  &  7.50e-02 \\
128  &   56.2123  &   32.2693  &    13.1  &   45  &        1  &    M3.25  &   16.29  &   14.60  &   12.33  &   11.49    &   11.13 	    &  1.27e+01  &  1.62e-01  &  8.85e+00  &  7.72e-02  &  6.04e+00  &  6.75e-02  &  3.65e+00  &  1.13e-01 \\
129  &   56.2240  &   32.1144  &     9.6  &   22  &        1  &       M4  &   18.04  &   16.03  &   13.33  &   12.37    &   11.96 	    &  6.04e+00  &  9.19e-02  &  4.42e+00  &  5.30e-02  &  3.16e+00  &  4.39e-02  &  1.52e+00  &  3.05e-02 \\
130  &   56.2317  &   32.1555  &     3.1  &   60  &        1  &       K4  &   16.02  &   14.39  &   11.85  &   10.75    &   10.36 	    &  2.41e+01  &  3.83e-01  &  1.77e+01  &  1.75e-01  &  1.16e+01  &  9.29e-02  &  6.66e+00  &  9.29e-02 \\
131  &   56.2338  &   32.0990  &     3.3  &   33  &        1  &    M2.75  &   16.50  &   14.95  &   13.07  &   12.27    &   11.99 	    &  5.57e+00  &  6.23e-02  &  3.54e+00  &  4.91e-02  &  2.54e+00  &  2.89e-02  &  1.42e+00  &  3.62e-02 \\
132  &   56.2339  &   32.1542  &     2.5  &   29  &        1  &       K0  &   14.69  &    5.26  &   11.02  &    9.99    &    9.47 	    &  6.62e+01  &  1.06e+00  &  6.28e+01  &  9.42e-01  &  4.42e+01  &  3.45e-01  &  3.81e+01  &  2.56e-01 \\
133  &   56.2563  &   32.1809  &     1.9  &   51  &      1,2  &       K0  &   14.26  &   13.26  &   11.86  &   11.13    &   10.88 	    &  1.36e+01  &  2.08e-01  &  9.33e+00  &  9.06e-02  &  6.29e+00  &  5.62e-02  &  3.77e+00  &  7.11e-02 \\
134  &   56.2573  &   32.2410  &    16.4  &   55  &      1,2  &       K4  &   14.80  &   13.46  &   11.34  &   10.40    &   10.04 	    &  3.18e+01  &  4.26e-01  &  1.91e+01  &  2.68e-01  &  1.45e+01  &  9.37e-02  &  8.61e+00  &  6.94e-02 \\
135  &   56.3166  &   32.5144  &     4.1  &   44  &        1  &     ---   &   15.17  &   13.90  &   11.96  &   11.07    &   10.77 	    &  1.59e+01  &  1.28e-01  &  1.05e+01  &  7.09e-02  &  7.04e+00  &  6.04e-02  &  3.78e+00  &  4.31e-02 \\
136  &   56.3250  &   32.3258  &     7.9  &   37  &        1  &     ---   &   17.72  &   15.87  &   13.35  &   12.40    &   12.05 	    &  6.12e+00  &  6.11e-02  &  3.96e+00  &  5.11e-02  &  2.55e+00  &  3.59e-02  &  1.49e+00  &  3.48e-02 \\
137  &   56.3353  &   32.1096  &     6.9  &   41  &        1  &       M1  &   14.90  &   13.42  &   11.12  &    9.98    &    9.35 	    &  7.93e+01  &  1.82e+00  &  8.07e+01  &  1.36e+00  &  7.68e+01  &  5.40e-01  &  7.31e+01  &  6.16e-01 \\
138  &   56.3371  &   32.3034  &     5.0  &   28  &        1  &     ---   &   18.43  &   16.67  &   15.59  &   14.70    &   14.46 	    &  5.55e-01  &  1.27e-02  &  3.86e-01  &  8.06e-03  &  2.38e-01  &  3.39e-02  &  1.65e-01  &  2.51e-02 \\
139  &   56.3378  &   32.3049  &     5.0  &   58  &        1  &     ---   &   17.70  &   15.72  &   12.87  &   11.82    &   11.40 	    &  1.17e+01  &  1.78e-01  &  7.51e+00  &  7.42e-02  &  5.26e+00  &  7.26e-02  &  2.81e+00  &  2.78e-02 \\
140  &   56.3846  &   32.0542  &     0.7  &   31  &        1  &       M3  &   16.03  &   14.60  &   12.89  &   12.14    &   11.89 	    &  5.40e+00  &  8.36e-02  &  3.88e+00  &  4.46e-02  &  2.58e+00  &  3.46e-02  &  1.48e+00  &  2.55e-02 \\
141  &   56.3980  &   31.9405  &     9.6  &   49  &        1  &     ---   &   16.64  &   15.08  &   12.76  &   11.79    &   11.41 	    &  9.91e+00  &  1.95e-01  &  6.12e+00  &  1.74e-01  &  3.87e+00  &  7.50e-02  &  2.16e+00  &  4.20e-02 \\
142  &   56.4327  &   32.4098  &    14.3  &   50  &        1  &     ---   &   17.62  &   15.82  &   13.42  &   12.53    &   12.17 	    &  4.97e+00  &  4.36e-02  &  3.34e+00  &  3.24e-02  &  2.36e+00  &  2.40e-02  &  1.37e+00  &  2.43e-02 \\
143  &   56.4448  &   32.4802  &    10.5  &   33  &        1  &     ---   &    ---   &    ---   &   12.77  &   11.79    &   11.41 	    &  8.56e+00  &  8.15e-02  &  6.04e+00  &  5.06e-02  &  4.62e+00  &  8.24e-02  &  2.49e+00  &  3.74e-02 \\
\enddata
\tablenotetext{1}{Power spectrum peak}
\tablenotetext{2}{Rotation period reported in this paper (1), Koziloglu et al. (2005) (2), Littlefair et al. (2005) (3), and Cohen et al. (2004) (4)}
\tablenotetext{3}{The $R_{C}$ and $I_{C}$ photometry of the following stars have been taken from Cohen et al. (2004), ID=45,49,79,82,87,96,111}
\end{deluxetable}

%%%%%%%%%%%%%%%%%%%%%%%%%%%%%%%%%%%%%%%%%%%%%%%%%%%
\clearpage
\pagestyle{plaintop}
\setlength{\voffset}{0mm}

\clearpage

\clearpage

\clearpage

\clearpage

\clearpage

\clearpage

\clearpage

\clearpage

\end{document}